\input harvmac
\noblackbox

\newcount\figno
\figno=0
\def\fig#1#2#3{
\par\begingroup\parindent=0pt\leftskip=1cm\rightskip=1cm\parindent=0pt
\baselineskip=11pt

\global\advance\figno by 1
\midinsert
\epsfxsize=#3
\centerline{\epsfbox{#2}}
\vskip 12pt
\centerline{{\bf Figure \the\figno:} #1}\par
\endinsert\endgroup\par}
\def\figlabel#1{\xdef#1{\the\figno}}

\def\risunok#1#2#3{\vskip 15pt 
\global\advance\figno by 1
\centerline{\epsfbox{#1}}
\vskip 10pt 
\centerline{{\bf Fig. #3: } #2}
\vskip 15pt }
\def\kartinka#1#2#3{
\par\begingroup\parindent=0pt\leftskip=1cm\rightskip=1cm\parindent=0pt
\baselineskip=11pt

\global\advance\figno by 1
\midinsert
\centerline{\epsfbox{#1}}
\vskip 12pt
\centerline{{\bf Fig. #3: }#2}\par
\endinsert\endgroup\par}

\def\np#1#2#3{Nucl. Phys. {\bf B#1} (#2) #3}
\def\pl#1#2#3{Phys. Lett. {\bf B#1} (#2) #3}

\def\physrev#1#2#3{Phys. Rev. {\bf D#1} (#2) #3}


\font\cmss=cmss10
\font\cmsss=cmss10 at 7pt
\def\rlx{\relax\leavevmode}
\def\inbar{\vrule height1.5ex width.4pt depth0pt}
\def\IC{\relax\,\hbox{$\inbar\kern-.3em{\rm C}$}}
\def\IN{\relax{\rm I\kern-.18em N}}
\def\IP{\relax{\rm I\kern-.18em P}}
\def\ZZ{\rlx\leavevmode\ifmmode\mathchoice{\hbox{\cmss Z\kern-.4em Z}}
  {\hbox{\cmss Z\kern-.4em Z}}{\lower.9pt\hbox{\cmsss Z\kern-.36em Z}}
  {\lower1.2pt\hbox{\cmsss Z\kern-.36em Z}}\else{\cmss Z\kern-.4em
  Z}\fi}
\def\IZ{\relax\ifmmode\mathchoice
{\hbox{\cmss Z\kern-.4em Z}}{\hbox{\cmss Z\kern-.4em Z}}
{\lower.9pt\hbox{\cmsss Z\kern-.4em Z}}
{\lower1.2pt\hbox{\cmsss Z\kern-.4em Z}}\else{\cmss Z\kern-.4em
Z}\fi}

\def\narrowplus{\kern -.04truein + \kern -.03truein}
\def\narrowminus{- \kern -.04truein}
\def\narrowminussub{\kern -.02truein - \kern -.01truein}

\def\half{{1\over 2}}

\def\w{{\omega}}

\def\frac#1#2{{#1\over #2}}

\def\Aut{O}
\def\vtr#1#2#3{\left[\matrix{#1\cr \matrix{#2 & #3}}\right]}
\def\vtrd#1#2#3#4#5{\left[\matrix{#1\cr \matrix{#2 & #3\cr
#4& #5}}\right]}
\def\vtrt#1#2#3#4#5#6#7{\left[\matrix{#1\cr
\matrix{#2& #3\cr #4& #5\cr #6& #7}}\right]}

\def\im{{\rm Im}\;}

\def\obr#1{{1\over #1}}

\def\Br{{\rm\bf B}}
\def\Cr{{\rm\bf C}}
\def\Dr{{\rm\bf D}}
\def\Fr{{\rm\bf F}}

\def\sym#1{{{\rm SYM}} _{#1 +1}}

\def\IZ{\relax\ifmmode\mathchoice
{\hbox{\cmss Z\kern-.4em Z}}{\hbox{\cmss Z\kern-.4em Z}}
{\lower.9pt\hbox{\cmsss Z\kern-.4em Z}}
{\lower1.2pt\hbox{\cmsss Z\kern-.4em Z}}\else{\cmss Z\kern-.4em
Z}\fi}
\def\IB{\relax{\rm I\kern-.18em B}}
\def\IC{{\relax\hbox{$\inbar\kern-.3em{\rm C}$}}}
\def\ID{\relax{\rm I\kern-.18em D}}
\def\IE{\relax{\rm I\kern-.18em E}}
\def\IF{\relax{\rm I\kern-.18em F}}
\def\IG{\relax\hbox{$\inbar\kern-.3em{\rm G}$}}
\def\IGa{\relax\hbox{${\rm I}\kern-.18em\Gamma$}}
\def\IH{\relax{\rm I\kern-.18em H}}
\def\II{\relax{\rm I\kern-.18em I}}
\def\IK{\relax{\rm I\kern-.18em K}}
\def\IP{\relax{\rm I\kern-.18em P}}
\def\IR{\relax{\rm I\kern-.18em R}}

\def\mod{\;{\rm mod}\;}

\font\cmss=cmss10 \font\cmsss=cmss10 at 7pt

\def\ge{\Gamma_8}  \def\gii{\Gamma_{1,1}}
 
\def\stw{\sqrt{2}} \def\stwi{\frac{1}{\sqrt{2}}}
\def\gsix{\Gamma_{(6)}}
\def\nik{{\cal N}}
%

%
%
\def\eqnn#1{\xdef #1{(\secsym\the\meqno)}\writedef{#1\leftbracket#1}%
\global\advance\meqno by1\wrlabeL#1}
\def\eqna#1{\xdef #1##1{\hbox{$(\secsym\the\meqno##1)$}}
\writedef{#1\numbersign1\leftbracket#1{\numbersign1}}%
\global\advance\meqno by1\wrlabeL{#1$\{\}$}}
\def\eqn#1#2{\xdef #1{(\secsym\the\meqno)}\writedef{#1\leftbracket#1}%
\global\advance\meqno by1$$#2\eqno#1\eqlabeL#1$$}

\input epsf

\lref\rchl{S.~Chaudhuri, G.~Hockney and J.~Lykken, 
``Maximally Supersymmetric String Theories in $D<10$'', hep-th/9505054}
\lref\rChP{S.~Chaudhuri, J.~Polchinski, ``Moduli Space of CHL String ``,
\physrev{52}{1995}{7168}, hep-th/9506048}
\lref\rDP{A.~Dabholkar and J.~Park, ``Strings on Orientifolds'', 
hep-th/9604178}
\lref\rSchwarzSen{J.H.~Schwarz, A.~Sen, ``Type IIA Dual of the
Six-Dimensional CHL Compactification'', 
\pl{357}{1995}{323}, hep-th/9507027}
\lref\rKachruKlemmOz{S.~Kachru, A.~Klemm and Y.~Oz, ``Calabi-Yau
Duals for CHL String'',
hep-th/9712035}
\lref\rWittenZt{E.~Witten, ``Toroidal Compactification
Without Vector Structure'', hep-th/9712028}
\lref\rWittenVD{E.~Witten, ``String Theory Dynamics in Various
Dimensions'', \np{443}{1995}{85-126}, hep-th/9503124}
\lref\rLL{K.~Landsteiner, E.~Lopez, ``New Curves from Branes'',
hep-th/9708118}
\lref\rPark{J.~Park, ``Orientifold and F-Theory Duals of CHL 
Strings'', hep-th/9611119}
\lref\rGSW{M.~Green, J.H.~Schwarz, E.~Witten, ``Superstring Theory''.}
\lref\rNarain{K.S.~Narain, \pl{169}{1985}{41}}
\lref\rNSW{K.S.~Narain, M.H.~Sarmadi and E.~Witten, ``A Note on Toroidal
Compactification of Heterotic String Theory'', 
\np{279}{1987}{369-379}}
\lref\rNSV{Narain, Sarmadi, Vafa, ``Asymmetric Orbifolds'',
\np{288}{1987}{551-577}}
\lref\rGPR{A.~Giveon, M.~Porrati, E.~Rabinovici, ``Target Space Duality
in String Theory'', hep-th/9401139}
\lref\rKac{V.G.~Kac, ``Infinite-Dimensional Lie algebras''}
\lref\rVinberg{A.L.~Onishchik, E.B.~Vinberg, 
``Lie Groups and Lie Algebras III'', Springer-Verlag 1994}
\lref\rConway{J.H.~Conway, N.J.A.~Sloane, ``Sphere Packings,
Lattices and Groups'', Springer-Verlag 1988}
\lref\rAspinwall{P.S.~Aspinwall, ``K3 Surfaces and String
Duality'', hep-th/9611137}
\lref\rLSMT{W.~Lerche, C.~Schweigert, R.~Minasian and S.~Theisen,
``A Note on the Geometry of CHL Heterotic Strings'', hep-th/9711104}
\lref\rS{C.~Schweigert, ``On moduli spaces of flat
connections with non-simply connected gauge group'', hep-th/9611092}
\lref\rSalamSezgin{A.~Salam, E.~Sezgin, ``Supergravities in Diverse
Dimensions'', Elsevier Science Publishers, B.V., and 
World Scientific Publishing Co., 1989}
\lref\rMorrison{D.R.~Morrison, ``On K3 surfaces with large Picard
number'', Invent.Math..~75, 105-121 (1984)}
\lref\rNikulinKt{V.V.~Nikulin, ``Finite Automorphism Groups of
Kahler $K3$ Surfaces'', Trans. Moscow Math. Soc. {\bf 38}, 71-135 (1980)}
\lref\rNikulin{V.V.~Nikulin, ``Integral symmetric bilinear forms and
some of their applications'', Math. USSR Izvestija {\bf 14}, 103-167
(1980)}
\lref\rGR{I.S.~Gradshteyn and I.M.~Ryzhik, ``Table of Integrals,
Series and Products'', Academic Press 1965}
\lref\rVafa{C.~Vafa, ``Evidence for F Theory'', \np{469}{1996}{403-418}}
\lref\rCasselman{W.A.~Casselman, ``Geometric Rationality
of Satake Compactifications'',$\;\;\;\;\;\;\;\;\;\;\;\;\;\;$
http://www.math.ubc.ca/people/faculty/cass/research.html}
\lref\rDouglasHull{M.R.~Douglas and C.~Hull, ``D-branes and
the Noncommutative Torus'', hep-th/9711165}
\lref\rGinspargCone{P.~Ginsparg, ``Curiosities at $c=1$'',
\np{295}{1988}{153-170}}
\lref\rGNO{
P.~Goddard, J.~Nyuts and D.~Olive, \np{125}{1977}{1}}
\lref\rSchwarzSenHet{J.H.~Schwarz, A.~Sen, ``Duality Symmetries
of 4D heterotic strings'', \pl{312}{1993}{105-114}}
\lref\rBPS{M.~Bershadsky, T.~Pantev and V.~Sadov, ``F-Theory and
Quantized Fluxes'', hep-th/9805056}
\lref\rIS{V.A.~Iskovskih, I.R.~Shafarevich, ``Algebraic Surfaces'',
in ``Algebraic Geometry - 2'', ed. I.R.~Shafarevich, 
Springer-Verlag 1996.}
\lref\rBKMT{P.~Berglund, A.~Klemm, P.~Mayr and S.~Theisen,
``On Type IIB Vacua With Varying Coupling Constant'', hep-th/9805189}

\Title{\vbox{\hbox{PUPT-1788}
\hbox{ITEP-TH-19/98}}}{Momentum Lattice for CHL String}
\smallskip
\centerline{
Andrei Mikhailov\footnote{$^\ast$}{On leave from the Institute
of Theoretical and Experimental Physics, Moscow, 117259, Russia.}
}
\medskip\centerline{\it Princeton University, Physics Department,
Princeton, NJ08544.}

\medskip\centerline{andrei@puhep1.princeton.edu}
\vskip 0.75in

We propose some analogue of the Narain lattice for CHL string.
The symmetries of this lattice are the symmetries
of the perturbative spectrum.  
We explain in this language the known results about the possible 
gauge groups in compactified theory. For the four-dimensional
theory, we explicitly describe the action of S-duality on 
the background fields. We show that the moduli spaces
of the six, seven and eight-dimensional compactifications
coincide with the moduli spaces of the conjectured
Type IIA, M Theory and F Theory duals. We classify the rational
components of the boundary
of the moduli space in  seven, eight and nine dimensions. 

\Date{}



\newsec{Introduction}

The nine-dimensional CHL string is an $N=1$
supersymmetric string theory which can be obtained as
an asymmetric orbifold of the ten-dimensional heterotic string.
This theory may be thought of as the compactification of the
$E_8\times E_8$ heterotic string on a circle $S^1$, with the exchange
of two $E_8$'s when one goes around  $S^1$. It is dual to
the compactification of $M$ theory on the M\"obius strip \rDP.
The CHL string was first discovered in eight dimensions in \rchl.
The authors of \rchl\ used a fermionic construction which
allowed them to study the theory near some special  points
in the moduli space.
In more invariant terms, the eight-dimensional theory can be
obtained as the compactification of the $Spin(32)/\IZ_2$ heterotic
string on a torus without vector structure \refs{\rWittenZt,\rLSMT}.

In dimensions less then nine, CHL string theory gives non-simply laced 
gauge groups
at special subsets of the moduli space \refs{\rchl,\rChP,\rPark}. 
In particular, in four dimensions the set of allowed gauge groups
 is self-dual. This is a manifestation
of the  $S$-duality in four-dimensional $N=4$ field theory.

The dual description in terms of Type IIA theory in dimensions 4, 5 and 6
was considered in \refs{\rSchwarzSen,\rKachruKlemmOz}. It involves
Calabi-Yau orbifolds with nontrivial RR $U(1)$ background turned on.
The description of the seven-dimensional compactification
in terms of $M$ theory on K3 surface with irremovable 
$D_4\oplus D_4$ singularity was obtained in the recent paper
\rWittenZt, using the results of \rLL. The eight-dimensional 
compactification was shown in \rWittenZt\ to be dual to the 
compactification of F Theory on K3 surface with irremovable
$D_8$ singularity.

In our paper, we study the perturbative spectrum of the CHL string
in nine and lower dimensions. We explicitly describe the T-duality group
as the group of symmetries of certain lattice. This lattice may be thought
of as an analogue of the Narain lattice for the CHL string. 
The moduli space of the theory is
\eqn\moduli{
\Aut(\Gamma_{(D)})\backslash O(18-D,10-D)/O(18-D)\times O(10-D)
}
where 
\eqn\general{
\Gamma_{(D)}=\Gamma_{9-D,9-D}(2)\oplus\gii\oplus\ge.
}

Here $\Gamma_{n,n}(2)$ means the lattice generated by $2n$ vectors
$\{e_i\}_{i=1,\ldots, n}$ and $\{f_i\}|_{i=1,\ldots,n}$ with the
scalar products $(e_i\cdot f_j)=2\delta_{ij}$. In general, given
the lattice $\Lambda$, we will denote $\Lambda(p)$ the lattice 
isomorphic to $\Lambda$ as an abelian group, but with the scalar
product multiplied by $p$ (it may be thought of as ``$\sqrt{p}\Lambda$'').
Our sign convention for the scalar product of the Narain lattice
is in agreement with \rNarain, and is opposite to the one usually
accepted in algebraic geometry \rAspinwall.

There are some useful equivalent forms for the lattice $\Gamma_{(D)}$. For
example,
\eqn\isomorphisms{\matrix{
\Gamma_{(8)}=\Gamma_{2,2}\oplus D_8\cr
\Gamma_{(7)}=\Gamma_{3,3}\oplus D_4\oplus D_4\cr
\Gamma_{(6)}=\Gamma_{4,4}\oplus D_8^*(2)
}}
where $D_8^*$ is the lattice, dual to $D_8$ (the weight lattice of $D_8$).

The lattice $\Gamma_{(D)}$ is not self-dual. 
In section 2, we discuss the perturbative spectrum of the CHL string
and the worldsheet current algebras.
In section 3, we construct the momentum lattice, prove that the
symmetries of this lattice are the symmetries of the perturbative
spectrum, and discuss the structure of the moduli space. 
In section 4, we study symmetry enhancements at the
special points of the moduli space, and explain how S-duality 
of the four-dimensional theory acts on the background fields. 
In section 5, we show that the moduli spaces we have found
coincide with what is expected from the known F-theory, M-theory, 
and Type IIA duals. 
Some useful results from the theory of lattices, relevant to our study,
are briefly reviewed in the Appendix A.
Section 6 and Appendices C and D 
are devoted to the study of the boundary of the moduli space.

Recently we have received the preprints \refs{\rBPS,\rBKMT}, 
where some problems discussed in our paper are studied from the different 
point of view.

\newsec{Perturbative Spectrum.}
To understand the structure of the perturbative spectrum,
let us first consider the nine-dimensional theory.
One way to construct it is to use the asymmetric orbifold of the heterotic 
string \rChP. Let us consider the $E_8\times E_8$ heterotic string 
compactified on a circle with the radius $r$ and orbifold the symmetry:
\eqn\orbifold{
x^9\to x^9+\pi r,\;\;\; x^I\to x^{I\pm 8}
}
where we have denoted $x^I$ the coordinates in the internal torus
${\bf R}^{16}/\Gamma_8\oplus\Gamma_8$. In our notations, capital letters
from the middle of the alphabet denote the directions along
the internal torus ${\bf R}^{16}/\Gamma_8\oplus\Gamma_8$, and
$(I\pm 8)$ means $(I+ 8)$ if $I<9$ or $(I-8)$ if $I\geq 9$. 
To understand the structure of twisted
and untwisted states, let us consider string diagrams with the topology
of a torus. For example, we can consider the scattering of
four or more gravitons with the momenta and polarizations
in the uncompactified directions.
We may consider a torus as a complex plane
with the identifications $z\sim z+1$ and $z\sim z+\tau$. In the 
${\bf Z}_2$-orbifold theory, we should consider the four possible 
boundary conditions on a torus: periodic in both directions,
periodic in $z\to z+1$ and periodic with the twist \orbifold\ 
in $z\to z+\tau$, periodic in $z\to z+\tau$ and with the twist in 
$z\to z+1$, and with the twists in both $z\to z+1$ and $z\to z+\tau$ 
directions. 
We will denote the corresponding contributions to the amplitude 
as $A_{++}(\tau)$, $A_{+-}(\tau)$, $A_{-+}(\tau)$ 
and $A_{--}(\tau)$, respectively. The sum
$$
A(\tau)=A_{++}(\tau)+A_{+-}(\tau)+A_{-+}(\tau)+A_{--}(\tau)
$$ 
is modular invariant. Consider first the expression 
$A_{++}(\tau)+A_{+-}(\tau)$. It may be calculated as the sum
of $e^{2\pi i(\tau L_0-\bar{\tau}\bar{L}_0)}$ over
those states of the heterotic string which are invariant
under $(-1)^{n_9}{\cal P}$, where $n_9$ is the momentum
along the ninth direction and $\cal P$ is the operator 
exchanging two $E_8$ indices. The explicit expressions 
are:
\eqn\Ap{
\matrix{
A_{++}(\tau)=\Phi(\{k_j\},\{\zeta_j\},\tau) 
{1\over qf(q)^{24}}(\im\tau)^{1\over 2}
\sum\limits_{m^9,n_9, I,J}q^{{1\over 2}p_L^2(m^9,n_9,I,J)}
\bar{q}^{{1\over 2}p_R^2(m^9,n_9,I,J)}\cr
A_{+-}(\tau)=\Phi(\{k_j\},\{\zeta_j\},\tau)
{1\over qf(q)^8f(q^2)^8}(\im\tau)^{1\over 2}
\sum\limits_{m^9,n_9, I}(-1)^{n_9}q^{{1\over 2}p_L^2(m^9,n_9,I,I)}
\bar{q}^{{1\over 2}p_R^2(m^9,n_9,I,I)}
}}
In these formulas, the following notations are used.
We have denoted 
$f(q)=\prod\limits_{k=1}^{\infty}(1-q^k)$ and $q=e^{2\pi i\tau}$. 
The momenta $p_L$ and $p_R$ are given by the formulae \rNSW:
\eqn\plpr{\eqalign{
p_L^9(m,n,P)=m+{g^{-1}\over 2}(n+A^I(P^I-{1\over 2}A^Im)) \cr
p_R^9(m,n,P)=-m+{g^{-1}\over 2}(n+A^I(P^I-{1\over 2}A^Im)) \cr
p_L^I(m,n,P)=P^I-A^Im}
} 
where $g=g_{99}$ is related to the radius of compactification
by the formula
\eqn\radius{
g=r^2=4r_{CHL}^2}
and $A$ is the Wilson line (for the orbifold projection to make
sense, the Wilson line should be diagonal).
$\Phi(\{k_j\},\{\zeta_j\},\tau)$ is some function of the momenta $k_j$ 
and polarizations $\zeta_j$ of the gravitons. The explicit expression
for this function may be found in \rGSW. The only thing we will need
to know about $\Phi$ is that it is a modular function of weight
$-4$:
\eqn\Phimodular{
\matrix{\Phi(\{k_j\},\{\zeta_j\},\tau+1)=
\Phi(\{k_j\},\{\zeta_j\},\tau),\cr
\Phi(\{k_j\},\{\zeta_j\},-{1\over \tau})=
\tau^4\Phi(\{k_j\},\{\zeta_j\},\tau)
}}

The expression $A_{-+}+A_{--}$ may be written as the summation
of $e^{2\pi i(\tau L_0-\bar{\tau}\bar{L}_0)}$ over the states 
from the twisted sector.
In the twisted sector, the string is closed only modulo the
symmetry \orbifold. This means, that the winding number along
the ninth direction, $m_9$, should be half-integer. As for the
internal coordinates, they have the following expansion in terms
of zero modes and oscillators:
\eqn\expansion{
\matrix{
X^I(\sigma+\tau)=x_0^I+p^I(\sigma+\tau)+
\sum\limits_{j\in{\bf Z}\backslash 0}
{\alpha^I_{j}\over j}e^{2\pi ij(\sigma+\tau)}+
\sum\limits_{j\in{\bf Z}}
{\alpha^I_{j+{1\over 2}}\over j+{1\over 2}}
e^{2\pi i(j+{1\over 2})(\sigma+\tau)}\cr
X^{I+8}(\sigma+\tau)=(x_0^I+p^I)+p^I(\sigma+\tau)+
\sum\limits_{j\in{\bf Z}\backslash 0}
{\alpha^I_{j}\over j}e^{2\pi ij(\sigma+\tau)}-
\sum\limits_{j\in{\bf Z}}
{\alpha^I_{j+{1\over 2}}\over j+{1\over 2}}e^{2\pi i(j+{1\over 2})
(\sigma+\tau)}
}}
and the eigenvalues of the operator $p^I$  belong to 
${1\over 2}\Gamma_8$. The oscillators
in the spatial direction and in the diagonal internal direction are
enumerated by integers, while the oscillators in the anti-diagonal direction
are enumerated by half-integers. This gives the normal ordering
constant $-{1\over 2}$ in the twisted sector\foot{
Oscillators with integer labels contribute $-{1\over 24}$ into
the normal ordering constant, while the oscillators with 
the half-integer labels contribute $+{1\over 48}$. This gives
the total central charge $-16\cdot{1\over 24}+8\cdot {1\over 48}=-\half$
in the twisted sector.},
which differs from the normal ordering constant $-1$ in the untwisted sector.
The explicit expression for $A_{-+}(\tau)$ is
\eqn\Amp{
\matrix{A_{-+}(\tau)=
{1\over \sqrt{q}f^8(q)f^8(\sqrt{q})}\;
(\im\tau)^{1\over 2}\;\Phi(\{k_j\},\{\zeta_j\},\tau)
\times\cr 
\cr
\times
\sum\limits_{\left[
\matrix{\scriptstyle
(m,n,P)\; :\;m\in\IZ+\half,\cr \scriptstyle n\in\IZ,\;\;
P\in\half\ge}\right]}q^{\half p_L^2(m,n,P,P)}
\bar{q}^{\half p_R^2(m,n,P,P)}
}}
and $A_{--}(\tau)=A_{-+}(\tau+1)$. There are half-integer
levels in the twisted sector, thus it is not true that 
$A_{-+}(\tau+1)=A_{-+}(\tau)$. The projector
\eqn\projector{
\half\left(1-(-1)^{n+P^2+2N'}\right)
}
where $P\in \ge\left(\obr{2}\right)$, plays the role of the orbifold
projection in the twisted sector.

The expression
\eqn\measure{
{d^2\tau\over (\im\tau)^2}(A_{++}(\tau)+A_{+-}(\tau)+
A_{-+}(\tau)+A_{--}(\tau))
}
is modular invariant. Indeed, it is clearly invariant under
$\tau\to\tau+1$. Also, the expression $A_{++}(\tau)$ is
invariant under $\tau\to -{1\over \tau}$, since it is 
just a heterotic string amplitude. Let us check that
\eqn\exchange{
A_{+-}(-1/\tau)=A_{-+}(\tau)
}
Indeed, it follows from the Poisson resummation formula that
\eqn\Poisson{
\matrix{(\im\tau)^{1\over 2}
\sum\limits_{m,n, I}(-1)^n q^{{1\over 2}p_L^2(m,n,I,I)}
\bar{q}^{{1\over 2}p_R^2(m,n,I,I)}=\cr =
{(\im(-1/\tau))^{1\over 2}\over 16\tau^4}
\sum\limits_{\matrix{\scriptstyle
(m,n,P)\; :\; m\in\IZ+\half,\cr \scriptstyle
n\in\IZ,\;\; P\in \half\ge}}
Q^{p_L^2(m,n,P,P)}\bar{Q}^{p_R^2(m,n,P,P)}
}}
where we have denoted $Q=e^{2\pi i (-1/\tau)}$. 
Also, we need the transformation law for the oscillator contributions:
\eqn\ofchl{
{1\over qf^8(q) f^8(q^2)}={16\tau^8\over
Q^{1/2}f^8(Q)f^8(\sqrt{Q})}
}

The transformation law \exchange\ follows from \Phimodular, 
\Poisson\ and \ofchl. Now, the invariance of $A_{--}$ under
$\tau\to -{1\over \tau}$ follows:
$$
A_{-+}(-1/\tau+1)=A_{+-}\left({\tau\over 1-\tau}\right)=
A_{+-}\left({1\over 1-\tau}\right)=A_{-+}(\tau-1)=A_{--}(\tau)
$$
This proves that the measure \measure\ is modular-invariant. 

Integration over the moduli space of the torus enforces
the level-matching condition, which is
\eqn\lmp{mn+\half(P^2+Q^2)+N-\tilde{N}-1=0}
in the untwisted sector and 
\eqn\lmt{mn+P^2+N-\tilde{N}-\half=0}
in the twisted sector (where $P\in\ge\left(\obr{2}\right)$). 

The mass formula reads
\eqn\mass{M^2={p_L^2\over 2}+{p_R^2\over 2}+N+\tilde{N}-a}
where $a$ is $1$ in the untwisted sector and $\half$ 
in the twisted sector.

Toroidal compactification to lower dimensions is straightforward.
There is one selected direction in the torus, which we will
call the ninth direction. The orbifold projection keeps
those states, which are invariant under the shift
by half a circle in this ninth direction with the exchange
of the coordinates in two internal $\IR^8/\ge$. In the twisted sector,
the winding number in the ninth direction is half-integer.

There are points in the moduli space where some states become
massless. The state in the twisted sector can become massless
if all the oscillators are in the ground state and the momentum
$K$ has $K^2=1$. In this case, we have an enhanced gauge
symmetry in space-time and Kac-Moody algebra on the worldsheet.
For the massless states in the twisted sector, some generators
of the worldsheet Kac-Moody algebra act between the twisted
and the untwisted sector. Indeed, we have the following set of
generators:
\eqn\KM{\matrix{
e_K(z)=\sqrt{2}V_K(z),\cr
h_K(z)=2iK\cdot\partial X(z),\cr
f_K(z)=\sqrt{2}V_{-K}(z)
}}
where $X^I(z)$ is a free left-moving boson with the operator
product expansion 
\eqn\XX{
\partial X^I(z)\partial X^J(0)=-{\delta_{IJ}\over z^2}+\ldots
}
and $V_K(z)$ is the vertex operator, creating a cut on the
worldsheet. It may be defined in terms of the path integral.
Insertion of such an operator means that we are integrating
over those fields which have a monodromy \orbifold\ when
we go around $z$, and
\eqn\VKz{
\oint_z \partial X=2\pi K
}
The conformal dimension of such an operator is $\half+\half K^2$.
We need $K^2=1$. Let us normalize $V_K(z)$ so that
\eqn\normalization{
\langle V_K(z)V_{-K}(0)\rangle ={1\over z^2}
}
The singular part of the operator product expansion of $\partial X$
and $V_{\pm K}(z)$ is determined by \VKz\ to be
\eqn\dXVK{
(K\cdot \partial X(z)) V_{\pm K}(0)=\mp{i\over z}V_{\pm K}(0)+\ldots
}
The three-point function is:
\eqn\threepoint{
\langle V_K(w) P\cdot \partial X(z) V_{-K}(0)\rangle = 
{i(P\cdot K)\over zw(w-z)}
}
--- indeed, this expression should be proportional to $1\over zw(w-z)$,
as follows from the conformal invariance, and the coefficient is
determined by \normalization\ and \dXVK. In this expression, $P$
is an arbitrary momentum from the twisted sector. From the formula 
for the three-point function and the normalization condition 
\normalization, we get the singular terms in $V_K(w)V_{-K}(0)$:
\eqn\VKVmK{
V_K(w)V_{-K}(0)={1\over w^2}+{i\over w}K\cdot\partial X(0)+\ldots
}
From the formulas \dXVK, \VKVmK\ and \XX, we get the operator product
expansions:
\eqn\KMope{\matrix{
h_K(z)e_K(0)={2\over z}e_K(0)+{\rm regular\;terms}\cr
e_K(z)f_K(0)={2\over z^2}+{1\over z}h_K(0)+{\rm regular\;terms}\cr
h_K(z)h_K(0)={4\over z^2}+{\rm regular\;terms}
}}
which is the operator product expansion for the generators
of the Kac-Moody algebra on level two.
We always get a level two algebra from the massless states
in the twisted sector. 

If the state in the untwisted sector becomes massless,
then, as we will see below, one can get both level two and 
level one current algebra, although it can be level one 
only in $D\leq 8$, not in the nine-dimensional theory. 

For example, consider the following state:
\eqn\exmassless{
|0,0,P,0>+|0,0,0,P>
}
where $P\in \Gamma_8$, $P^2=2$. If this state is massless, 
we have the following currents:
\eqn\untwAKM{\matrix{
e_P(z)=:e^{iP\cdot X^{(1)}}:+:e^{iP\cdot X^{(2)}}:,\cr
h_P(z)=iP(\partial X^{(1)}+\partial X^{(2)}),\cr
f_P(z)=:e^{-iP\cdot X^{(1)}}:+:e^{-iP\cdot X^{(2)}}:
}}
--- here we use the notation $X^{(1)}$ and $X^{(2)}$ for the
components of $X$ in the first and the second torus in
$$
{\bf R}^{16}/\Gamma_8\oplus\Gamma_8=
({\bf R}^8/\Gamma_8)\times ({\bf R}^8/\Gamma_8)
$$

We may compare \untwAKM\ to the Kac-Moody generators for the 
level one algebra on the worldsheet of the usual 
heterotic string:
\eqn\usualAKM{\matrix{
e_P(z)=:e^{iP\cdot X}:,\cr
h_P(z)=iP\cdot\partial X,\cr
f_P(z)=:e^{-iP\cdot X}:
}}

The $\widehat{sl(2)}$ algebra generated by \untwAKM\ is
on the level two.
If we compactify to some dimension lower then nine, we can
get some level one algebras (corresponding to the massless states
from the untwisted sector). For example, we may compactify the
nine-dimensional CHL string on the circle of the self-dual radius. 
The $sl(2)$ current algebra corresponding to the massless
states with winding and momentum along this circle may be constructed
exactly in the same way, as for the usual heterotic string. It is
on level one.
 
\newsec{The Lattice.}
\subsec{Momentum of a State.}
We start with the nine-dimensional theory.
Let us associate with each state a vector from the lattice
$$\Gamma_{1,1}\oplus \Gamma_8$$
This is done as follows. To the untwisted state
with the internal momentum $(P,Q)$, winding number $m$ and momentum
$n$ along the ninth direction,  
we associate the vector of the form $(l,n,R)$, where
$$\matrix{R=P+Q\cr l=2m}$$
To the twisted state which has the internal momentum 
$({P\over 2},{P\over 2})$, and the momentum and winding
number $(n,m)$ in the ninth direction, we associate the 
lattice vector $(l,n,P)$, with $l=2m$.
The left and right components $p_L^9$ and $p_R^9$ of the momentum,
as well as the diagonal part of the internal momentum $p_L^I$,
are, in the given background,  functions of $(l,n,R)$ only;
thus the dependence of the mass of
a given state on the background fields is only through the
coupling of the background to the vector $(l,n,R)$. This is one of the
motivations for calling $(l,n,R)$ the momentum. The other reason is that,
as we will see later, the perturbative spectrum is invariant under 
those transformations of the background which can be interpreted
as the symmetries of the lattice generated by the momenta $(l,n,R)$. 

The mass formula in both twisted and untwisted sectors,
with the diagonal Wilson line $A=(a,a)$ turned on, reads as:
\eqn\mass{
M^2(l,n,P,Q)={1\over 4}(R-la)^2+{g\over 4}l^2+
{1\over 4g}\left(n+aR-
{l\over 2}a^2\right)^2+N'+\tilde{N}-1
}
where the modified oscillator number 
$N'$ is defined as $N'=N+\half$ in the twisted sector, and 
as $N'=N+{(P-Q)^2\over 4}$ in the untwisted sector, $N$ being the
oscillator number. In both sectors, $N'$ may be integer or half-integer.
Our notations are summarized in the table:
\vskip 10 pt
$$
\def\tbntry#1{\vbox to 23 pt{\vfill \hbox{#1}\vfill }}
\hbox{\vrule width 1dd
      \vbox{\hrule height 1dd 
            \hbox{\vrule
                  \hbox to 100 pt{
                  \hfill\tbntry{Sector}\hfill }
                  \vrule width 1dd
                  \hbox to 100 pt{
                  \hfill\tbntry{Untwisted}\hfill }
                  \vrule
                  \hbox to 100 pt{
                  \hfill\tbntry{Twisted}\hfill }
                  \vrule width 1dd
                 }
            \hrule 
            \hbox{\vrule
                  \hbox to 100 pt{
                  \hfill\tbntry{Momentum}\hfill }
                  \vrule width 1dd
                  \hbox to 100 pt{
                  \hfill\tbntry{$(m,n,P,Q)$}\hfill }
                  \vrule
                  \hbox to 100 pt{
                  \hfill\tbntry{$(m+{1\over 2},n,
                                  {1\over 2}P,{1\over 2} P)$}\hfill }
                  \vrule width 1dd
                 }
            \hrule 
            \hbox{\vrule
                  \hbox to 100 pt{
                  \hfill\tbntry{$(l,n,R)$}\hfill }
                  \vrule width 1dd
                  \hbox to 100 pt{
                  \hfill\tbntry{$(2m,n,P+Q)$}\hfill }
                  \vrule
                  \hbox to 100 pt{
                  \hfill\tbntry{$(2m+1,n,P)$}\hfill }
                  \vrule width 1dd
                 }
            \hrule 
            \hbox{\vrule
                  \hbox to 100 pt{
                  \hfill\tbntry{$N'$}\hfill }
                  \vrule width 1dd
                  \hbox to 100 pt{
                  \hfill\tbntry{$N+{(P-Q)^2\over 4}$}\hfill }
                  \vrule
                  \hbox to 100 pt{
                  \hfill\tbntry{$N+{1\over 2}$}\hfill }
                  \vrule width 1dd
                 }  
            \hrule height 1dd
         }
     }
$$
\hskip 10pt
For each vector from the lattice, we have the whole infinite tower
of states associated to it. The structure of this tower is different
for different vectors. We will prove that the symmetries of the
lattice are the symmetries of the perturbative spectrum. It turns out,
that the vectors in the momentum lattice may be naturally 
grouped into three classes invariant under the symmetry group
of the lattice, so that the spectrum of the massive states 
with given momentum depends only on which class the momentum
belongs to.
To prove this, let us introduce the generating function
for the number of states with given momentum. Since the oscillators
in the spatial direction are completely decoupled, we can without
any loss of generality consider only those states for which 
these spatial oscillators are not excited. We define the generating function
for the number of states with given momentum $(l,n,R)$ as:
\eqn\generating{
F(q)=\sum d(N',(l,n,R))q^{N'}
}
where $d(N')$ is the number of states with the given value of $N'$
and given momentum $(l,n,R)$. We get the following expressions
for the generating function, depending on the momentum $(l,n,R)$:

{\bf A.} 
$l\in 2{\bf Z},\;\; R\in 2\Gamma_8$ (untwisted sector, internal momentum
divisible by two):
\eqn\first{
F_1(q,n \mod 2)={1\over 2}\left({\Theta_8(2\tau)\over f^{24}(q)}+
{(-1)^n\over f^8(q)f^8(q^2)}\right) 
}
--- this expression depends on whether the momentum $n$ along the circle 
is odd or even, because the orbifold projection involves $(-1)^n$.

{\bf B.} $l\in 2{\bf Z},\;\; R\in \Gamma_8\backslash 2\Gamma_8$ 
(untwisted sector, internal momentum cannot be divided by two):
\eqn\second{
F_2(q,\bar{R})=
{e^{{\pi i\tau\over 2}R^2}\over 2f^{24}(q)}\Theta_8(2\tau|\tau R)=
{1\over 2f^{24}(q)}\sum\limits_{Q\in \Gamma_8} e^{2\pi i\tau (Q+R/2)^2}
}
--- this expression depends only on the conjugacy class
of $R$ in $\Gamma_8/2\Gamma_8$. 

{\bf C.} $l\in 2{\bf Z}+1,\;\;n\in {\bf Z}$ (twisted sector):
\eqn\third{
F_3(q, \half R^2 + n\; \mod 2)=
{\sqrt{q}\over 2 f^8(q)}\left[{1\over f^8(\sqrt{q})}-
{(-1)^{R^2/2+n}\over f^8(-\sqrt{q})}\right]
}
In these formulae we have used the following notation
for the $\theta$-function of the lattice $\Gamma_8$: 
\eqn\theta{
\Theta_8(\tau|z)=\sum\limits_{S\in\ge} 
e^{\pi i \tau S^2+2\pi i (S\cdot z)}
}
We will explain momentarily how to derive these formulae.
But first let us notice, that \first---\third\ give six 
different expressions for
generating function, depending on whether $l$ and $n$ are even
or odd, and on the conjugacy class of the momentum $R$ modulo $2\ge$.
If the generating functions in all these situations were different, 
the spectrum would not
be invariant under the symmetries of the lattice. For example,
the symmetry $(l,n,R)\to (n,l,R)$ for even $l$ and odd $n$
exchanges the states from the untwisted sector with the states
from the twisted sector. Since the allowed momenta and
oscillator numbers are completely different in these two sectors,
it is not obvious that there exists a symmetry between them.
Fortunately, there is such a symmetry\foot{There are many examples
in string theory, where different sets of creation-annihilation
operators give the same spectrum. Correspondence between bosons
and fermions is one of them. An example of the symmetry mixing twisted
and untwisted states may be found in \rGinspargCone}.
In fact, as we will see later,
the functions $F_1$, $F_2$ and $F_3$ satisfy certain identities,
which reduce the number of different classes of vectors from
six to three. 

Let us explain how to derive \first---\third. To get \first,
we notice that those states from the untwisted sector which have
$P\neq Q$ may be considered as the states of the heterotic string.
The internal momenta $(P,Q)$ and $(Q,P)$ which differ only in the
order of two components, give the same untwisted state after the
orbifold projection (which gives a factor of $\half$ on
the right hand side). For $R\in 2\Gamma_8$, we may write $R=P+Q$ with
$P=\half R+S\in\ge$ and $Q=\half R-S\in\ge$. The momentum square
$\half P^2+\half Q^2$ contains 
${S^2\over 2}+{S^2\over 2}=S^2$ giving a contribution to $N'$ in the mass
formula. After performing 
a summation over $S\in\ge$ and taking into account the oscillator
contributions, we get the first term on the right hand side
of \first. To get the second term, we have
to take into account that for the diagonal momentum $P=Q$
(or $S=0$) not all the excited levels of the oscillators are
allowed by the orbifold projection, but only those which
have $n$ plus the number of antidiagonal creation operators
even. 

To get \second, we notice that only the states with
$P\neq Q$ may correspond to $R\in\ge\backslash 2\ge$ and $n$ even. 
For each such state, the momentum may be written as $(Q+R,-Q)$, 
which gives the contribution $(Q+R/2)^2$ to $N'$. 

In the formula \third, the first term on the right hand side
comes from $A_{-+}(\tau)$, and the second term 
comes from $A_{--}(\tau)$. In this case, the term $N'$ in the
mass formula gets contributions from the oscillators only.

It turns out that the spectrum of massive states actually depends on
which of the following classes the point of the lattice belongs to:

{\bf 1.} $(l,n,R)\in 2(\gii\oplus\ge)$,

{\bf 2.} $ln+\half R^2$ is even, but $(l,n,R)$ is not in the previous
class,

{\bf 3.} $ln+\half R^2$ is odd.

{\bf Proof.} Begin with the class {\bf 3}.
The vectors from this class correspond to either case
{\bf B} with odd $\half R^2$, or to the case {\bf C}
with odd $\half R^2+n$. Notice that if $\half R^2$ is odd,
then one can always find such a vector $Q\in 2\Gamma_8$
that $R=P+Q$ and $P\in\Delta$ --- the root of $E_8$.
Indeed, consider the vector $\half R$. It is known \rConway,
that the covering radius of $\ge$ is $1$. This means that 
there is some vector of the lattice within the distance $1$ to $\half R$. 
Let us call it $Q$. Then, we have:
$$ (\half R-Q)^2={1\over 2} $$
--- indeed, the right hand side has to be half-integer and less
or equal then one. 
Thus, we have to prove that
\eqn\Proot{F_2(q,P)=F_3(q, 1)}
for $P$ a root. 
Let us prove a slightly more general identity:
\eqn\iiplusiiii{
\sum\limits_{Q\in\Gamma_8}
\left(e^{2\pi i\tau(Q+R_1/2)^2}+
e^{2\pi i\tau(Q+R_2/2)^2}\right)=2\sqrt{q}{f^{16}(q)\over f^8(\sqrt{q})}
}
where $R_1=(1,1,0^{(6)})$ and $R_2=(1^{(4)},0^{(4)})$.
This may be shown directly, using the known formulae for theta-series
(see, for example, \rGR, formulas 8.180 and 8.181). We will actually need
the following identities:
\eqn\thetas{\matrix{
\sum\limits_{n\in\IZ}e^{\pi i\tau n^2}=
\prod\limits_{n=1}^{\infty}(1+q^{n-\half})^2(1-q^n),\cr
\sum\limits_{n\in\IZ}e^{\pi i\tau \left( n+\half\right)^2}=
2q^{1\over 8}\prod\limits_{n=1}^{\infty}(1-q^{2n})(1+q^n)
}}
Using these two identities, we get:
$$\matrix{
\sum\limits_{n_1+\ldots+n_8\;\;{\rm even}}&
\left(e^{2\pi i\tau\left(\left({1\over 2}+n_1\right)^2+
\left({1\over 2}+n_2\right)^2+\sum_{j=3}^8 n_j^2\right)}\right.
+&\hskip -8pt \left.e^{2\pi i\tau\left(n_1^2+n_2^2+
\sum_{j=3}^8\left({1\over 2}+n_j\right)^2\right)}+\right.
\cr
 &\left.+
2e^{2\pi i\tau\left(\sum_{j=1}^4\left({1\over 2}+n_j\right)^2+
\sum_{j=5}^8n_j^2\right)}\right)=& }
$$
$$
={1\over 2}\sum\limits_{n_1,n_2,n_3,n_4}
e^{2\pi i\tau(n_1^2+n_2^2)}e^{2\pi i\tau\left(
\left({1\over 2}+n_3\right)^2+\left({1\over 2}+n_4\right)^2\right)}
\left[\sum\limits_{n_+,n_-}e^{\pi i\tau(n_+^2+n_-^2)}\right]^2=
2\sqrt{q}{f^{16}(q)\over f^8(\sqrt{q})}
$$
The part of this expression odd under $\sqrt{q}\to -\sqrt{q}$
gives the required identity \Proot.

Now let us consider the class {\bf 2}. The vectors of this 
class come from either type {\bf A} with odd $n$, or 
type {\bf B} with $\half R^2$ even, or type {\bf C}
with $\half R^2+n$ even. Notice that if $\half R^2$ even but $R$ is
not in $2\ge$, then there exists such a vector $S\in\ge$ that 
$S^2=4$ and $R\equiv S \mod 2\ge$. Also, all the vectors in 
$\ge$ with length square $4$ are equivalent modulo Weyl group.
Thus, we have to prove that
\eqn\typetwoI{ F_2(q,S)=F_3(q,0)}
and
\eqn\typetwoII{ F_1(q,1)=F_2(q,S)}
where $S=(2,0^{(7)})$ (or $S=(1^{(4)},0^{(4)})$ --- since it is
related to $(2,0^{(7)})$ via the Weyl group, we get the 
same $F_2(q,S)$). The identity \typetwoI\ is the part
of \iiplusiiii, even under $\sqrt{q}\to -\sqrt{q}$.  
To prove \typetwoII, we write
\eqn\te{
\Theta_8(2\tau)=\half\left[{f^{16}(-q)\over f^8(q^2)}+
{f^{16}(q)\over f^8(q^2)}\right]+128q^2{f^{16}(q^4)\over f^8(q^2)}
}
--- this expression may be derived from the explicit form
of the $E_8$ lattice, using  \thetas, or from the fermionic 
description of the heterotic string \rGSW. Then, we get:
\eqn\vc{
\Theta_8(2\tau)-{f^{16}(q)\over f^8(q^2)}=
\half\left[{f^{16}(-q)\over f^8(q^2)}-
{f^{16}(q)\over f^8(q^2)}\right]+128q^2{f^{16}(q^4)\over f^8(q^2)}
}
Notice that the first term on the right hand side is
the sum of $e^{2\pi i\tau P^2}$
over $P$ in the vector conjugacy class of $so(16)$, while
the second one is the sum over $P$ in the spinor conjugacy class.
This is exactly how we get $f^{24}(q)F_2(q, (2,0^{(7)}))$.

As for the vectors of class {\bf 1}, they appear only in type {\bf A},
and it is evident that the spectrum of the massive states is the same
for all of them. {\bf This completes the proof.}

Since the definition of the classes {\bf 1,2} and {\bf 3} is given
in an invariant way, any automorphism of the lattice 
$\Gamma_{1,1}\oplus\ge$ is actually a symmetry of the perturbative 
spectrum.

If we further compactify CHL string on a $d$-dimensional torus,
we get the lattice
\eqn\lattice{\Gamma_{(9-d)}=\Gamma_{d,d}(2)\oplus\gii\oplus\ge}
The coefficient $2$ in the scalar product for 
$\Gamma_{d,d}(2)$ appears
because the momentum $m$ along any direction in ${\bf T}^d$ should
be integer, which gives even $l$. The lattice consisting
of the vectors of the form $(2m,n),\;\;m,n\in\IZ$ with the
length square $||(2m,n)||^2=4mn$ is $\gii(2)$. 
For the orbifold projection to make sense, Wilson lines in
all the compactified directions should be diagonal. 

As in the nine-dimensional theory, 
the structure of the massive states depends on the class
of the vector. The sublattice of the vectors of class {\bf 1}
is now defined as 
\eqn\sublattice{\Gamma_{d,d}(2)\oplus 2(\gii\oplus\ge)
\subset\Gamma
}
This sublattice can be described in an invariant way as the
sublattice of vectors of even level\foot{The level of the lattice
vector is the greatest common divisor of all the scalar products
of this vector with the other vectors in the lattice.}
 in $\Gamma$. The other two classes are defined in the same way
as before. This classification is preserved by all the symmetries
of the lattice. Thus, $\Aut(\Gamma)$ is a group of symmetries
of the perturbative spectrum. 

Given the lattice 
$\Gamma_{(D)}=\Gamma_{d,d}(2)\oplus \Gamma_{1,1}\oplus\ge$, consider 
the dual lattice:
\eqn\DualLattice{
(\Gamma_{d,d}(2)\oplus \Gamma_{1,1}\oplus\ge)^*
\simeq
\Gamma_{d,d}\left(\half\right) \oplus \Gamma_{1,1}\oplus\ge
}
For an arbitrary lattice $L$, we have 
\eqn\AutsGen{
\Aut(L)\simeq\Aut(L^*) 
}
(indeed, each symmetry of $L$ gives the symmetry of $L^*$,
and vice versa: it follows from $(L^*)^*=L$, that the symmetry
of $L^*$  corresponds to some symmetry of $L$). Since the lattice
$L(2)$ may be obtained from the lattice $L$ by rescaling
with the factor $\stw$, the groups $\Aut(L)$ and $\Aut(L(2))$ 
are isomorphic. Taking this into account, we derive from 
\DualLattice\ and \AutsGen, that 
\eqn\Auts{
\Aut(\Gamma_{d,d}(2)\oplus \Gamma_{1,1}\oplus\ge)\simeq
\Aut(\Gamma_{d,d}\oplus\Gamma_{1,1}(2)\oplus\ge(2))
}
Notice that $\ge(2)$ can be embedded as the diagonal
sublattice into $\ge\oplus\ge$. Then, the
isomorphism \Auts\ shows, in particular, that those symmetries
of the $10-d$-dimensional heterotic spectrum, which preserve
the diagonal Wilson lines, are also the symmetries of the CHL
spectrum. Of course, this is in agreement with what one would expect.

\subsec{The Moduli Space.}

Let us show that the moduli space of the compactified CHL string is
the Grassmanian manifold of planes with signature
$(0,d+9)$ in the space $\IR^{d+1,d+9}$, modulo the symmetries
of the lattice:
\eqn\GrDef{\matrix{
{\cal M}_{9-d}=O(\Gamma_{(9-d)})\backslash Gr((0,d+9),\IR^{d+1,d+9})=
O(\Gamma_{(9-d)})\backslash Gr((d+1,0),\IR^{d+1,d+9})
=\cr \cr 
=O(\Gamma_{(9-d)})\backslash O(d+1,d+9)/O(d+1)\times O(d+9)
}}

The choice of the 
background may be thought of as giving the decomposition of the momenta
in the lattice into the left- and the right-moving components, according
to the equation:
\eqn\plprDiag{
\matrix{
p_L^i(l,n,R)={1\over 2}\left[ l^i+
E^{ij}(n_j+a_j^I(R^I-{1\over 2}a_k^I l^k))\right]\cr
p_R^i(l,n,R)={1\over 2}\left[-l^i+
E^{ij}(n_j+a_j^I(R^I-{1\over 2}a_k^I l^k))\right]\cr
p_L^I=R^I-a^I_i l^i,\;\; i=1,\ldots, 8
}}
--- here we have introduced, following \rGPR,
the matrix $E_{ij}=g_{ij}+b_{ij}$.  
For the background to have geometrical meaning, the metric $g_{ij}$
should be nonnegative definite: $g_{ij}v^iv^j\geq 0$ for any vector $v$.
Then, the inverse matrix $E^{ij}$ is well defined  if and only if
$g_{ij}$ is nondegenerate. 

Let us consider the plane in $\IR\otimes\Gamma_{(D)}$
specified by the equation\foot{The tensor product
of the lattice with $\IR$ is the linear space, generated
by the triples $(l,n,R)$ with $l,n$ real (not necessarily integer)
numbers, and $R$ arbitrary vector in $\IR^8$ (not
necessarily belonging to $\Gamma_8\subset\IR^8$). The
system of $d+1$ equations $p_R=0$ gives some $d+9$-dimensional
plane in this vector space.}  $p_R=0$. It is clear,
that different backgrounds will give us different planes. Thus, we have
a map from the space of backgrounds to the space of $(0,d+9)$-planes 
in $\IR^{d+1,d+9}$. This map is well defined everywhere except for those
points in the moduli space, where the metric $g_{ij}$ is degenerate.

Let us consider the inverse map. Suppose that the $(0,d+9)$-plane
in $\IR^{d+1,d+9}$ is given by the equation 
\eqn\inverse{
G_{ij}l^j=n_i+(a_i\cdot R)
}
Then, the corresponding background is specified as follows:
$a_i$ is the Wilson line, and $G_{ij}=E_{ij}+\half (a_i\cdot a_j)$.
Notice that $a_i$ and $G_{ij}$ are well defined for all planes
except for those which contain vectors with zero $l$- and $R$-
components (and nonzero $n$-components). This means, that the 
corresponding direction on the plane is lightlike. Such planes are
on the boundary of the moduli space. 

We have shown that the points of the moduli space may be
parametrized by the planes in $\IR^{d+1,d+9}$. 
The metric on this moduli space is read from the low-energy
effective action, which is fixed by supersymmetry. The field
content is that of the compactification of 10-dimensional
$N=1$ supergravity interacting with 8 vector fields.
The moduli space for these theories is locally \rSalamSezgin:
\eqn\sugra{
\frac{SO(10-D,18-D)}{SO(10-D)\times SO(18-D)}\times \IR
}
for $D\geq 5$ (the factor $\IR$ specifies the dilaton expectation
value) and
\eqn\sugrafive{
\frac{SO(6,14)}{SO(6)\times SO(14)}\times {SU(1,1)\over U(1)}
}
in $D=4$ (the last factor specifies the expectation values
of dilaton and axion, which is a dual of $B$ field).

Globally, the moduli space differs from \sugra\ and \sugrafive, 
because we have to
take into account that some apparently different backgrounds 
should be in fact identified because of the dualities. 
We certainly should identify those backgrounds which are
related by series of reflections, since reflections are in fact
gauge symmetries \rGPR.
For the lattice $\ge\oplus\gii$, it is known 
\rConway, that the whole symmetry group is in fact
generated by the reflections. We do not know if the same result
is true for the non-Lorentzian lattices, which arise when
we study compactifications to $D<9$. Nevertheless, we assume that
we should identify those backgrounds, which may be related
by the symmetry of the momentum lattice. This leads to the
following form of the moduli space for $9-d$-dimensional
theory:
\eqn\Mcal{
{\cal M}_{9-d}=
O(\Gamma_{1,1}\oplus\Gamma_{d,d}(2)\oplus\ge)
\backslash O(d+1,d+9)/O(d+1)\times O(d+9)
}

\subsec{Some useful isomorphisms.}

We will need some facts about the lattice  
$$
\Lambda_d=\Gamma_{d,d}(2)\oplus\Gamma_8
$$
This lattice may be defined as the sublattice of 
$II_{d,8+d}=\Gamma_{d,d}\oplus\Gamma_8$, consisting of those
vectors whose scalar product with any vector from a
rational light-like plane of maximal dimension\foot{There
is only one such plane, modulo automorphisms of $II_{d,d+8}$.
This may be proven as follows. Given a light-like plane $\omega$,
consider some primitive vector $v\in\omega$. Since the lattice is self-dual,
one can find a primitive lightlike vector $v^*$, such that 
$(v^*\cdot v)=1$. Consider the sublattice orthogonal
to $v$ and $v^*$. This sublattice is even and self-dual, thus
it is isomorphic to $II_{d-1,d-1+8}$. The intersection of $\omega$
with this plane is a rational lightlike plane in $II_{d-1,d-1+8}$.
After applying this procedure $d$ times, we find that $\omega$
is the standard light-like plane in 
$\Gamma_{d,d}\subset \Gamma_{d,d}\oplus \ge$.}
is even. Indeed, let us take the lightlike plane consisting
of the vectors of the form 
\eqn\lightlike{
\vtrt{0}{*}{0}{\ldots}{\ldots}{*}{0}
}
(we write elements of $\Gamma_{d,d}\oplus\ge$ as
$$
\vtrt{R}{l_1}{n_1}{\ldots}{\ldots}{l_d}{n_d}
$$
and the length square is $2l_1 n_1+\ldots +2l_d n_d+ P^2$)

Then, those vectors which have even scalar products with 
the vectors from the plane \lightlike , are of the form
$$
\vtrt{P}{2m_1}{n_1}{\ldots}{\ldots}{2m_d}{n_d}
$$
with integer $m$'s, and form a sublattice $\Lambda_d$.

Since all the maximal rational lightlike planes are equivalent,
we may have some more convenient choices. For example, to
get $\Lambda_1$, we may use the following lightlike vector:
\eqn\llv{
\vtr{\alpha_0+\alpha_2+\alpha_4+\alpha_6}{2}{-2}
}
Here we use the following enumeration of the roots of $\ge$:

\kartinka{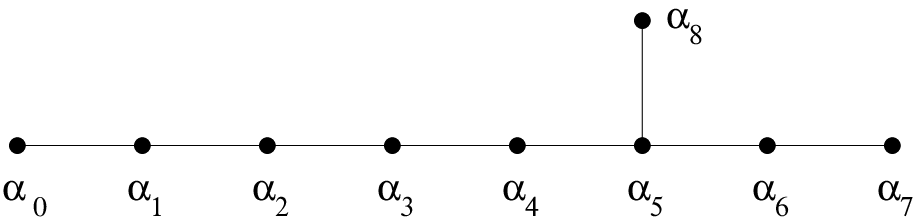}{Root system of $\ge$.}{1}

This is the extended Dynkin diagram; the roots are linearly dependent:
$$
\alpha_0+2\alpha_1+3\alpha_2+4\alpha_3+5\alpha_4+6\alpha_5+4\alpha_6+
2\alpha_7+3\alpha_8=0
$$

The sublattice of those vectors in $\ge$ which have even scalar 
product with $\alpha_0+\alpha_2+\alpha_4+\alpha_6$ is isomorphic
to $D_8$ (the lattice $D_n$ is defined as $\{(m_1,\ldots, m_n)|
\sum m_j\in 2\IZ\}$).
Thus, we get
\eqn\lambdaone{
\Lambda_1\simeq \gii\oplus D_8
}
This means, that
$$
\Gamma_{(8)}\simeq \Gamma_{2,2}\oplus D_8
$$

Let us consider the lattice
\eqn\stwde{
\gii(2)\oplus D_8
}
This lattice may be considered as the sublattice of those
vectors in $\gii\oplus D_8$, which have even scalar
product with the vector $v=\vtr{0}{0}{1}$. Let $\{u_1,\ldots,u_8\}$
be the orthonormal basis in $\IR^8$, then the lattice
$D_8$ consists of the linear combinations $\sum x_i u_i$
with integer coefficients $x_i$, subject
to the condition $\sum x_i\in 2\IZ$. Let us consider
a composition of the reflection in the vector
$\vtr{u_1-u_3}{1}{0}$ and $\vtr{u_3-u_2+u_5-u_6}{-1}{1}$.
This symmetry transforms our vector $v$ to the vector
\eqn\VectorV{
\vtr{-u_1-u_2+u_5-u_6+2u_3}{-2}{2}
}
and the vectors in $\gii\oplus D_8$ having even
scalar product with this vector form the sublattice
$\gii\oplus D_4\oplus D_4$. 

This gives us the following
isomorphisms:
\eqn\dfdf{\matrix{
\gii(2)\oplus D_8\simeq \gii\oplus D_4\oplus D_4,\cr
\Lambda_2\simeq \Gamma_{2,2}\oplus D_4\oplus D_4
}}

Thus, the momentum lattice for the seven-dimensional
compactification is
\eqn\sevenD{
\Gamma_{(7)}\simeq \Gamma_{3,3}\oplus D_4\oplus D_4
}

We will need also the following isomorphism:
\eqn\PDe{
\gii(2)\oplus D_4\oplus D_4\simeq 
\gii\oplus D_8^*(2)
}
where $D_8^*$ is the weight lattice of $D_8$; it consists
of the linear combinations $\sum x_iu_i$ with the
restriction that all $x_i$ are either simultaneously
integer, or simultaneously half-integer. We can prove it in
the same way as we had proven \dfdf. 
The lattice $\Gamma_{1,1}\oplus D_8^*(-2)$ is
the sublattice of $\gii\oplus D_4\oplus D_4$, consisting of those vectors
which have even scalar product with the vector  \VectorV: 
that this vector actually belongs to 
$\gii\oplus D_4\oplus D_4\subset\gii\oplus D_8$, and the symmetries,
relating this vector to $\vtr{0}{0}{1}$, are
the symmetries of $\gii\oplus D_4\oplus D_4$.
The isomorphism \PDe\ implies that the momentum lattice of the 
six-dimensional theory may be represented as:
\eqn\sixD{
\Gamma_{(6)}\simeq \Gamma_{4,4}\oplus D_8^*(2)
}

Alternative proofs of these identities using the technique of
discriminant-forms are given in Appendix A.

\newsec{Gauge groups in the compactified theory.}

In this section we will show that the results of S.~Chaudhuri
and J.~Polchinski \rChP\ about the possible gauge groups 
follow from our description of the momentum lattice. 
We will find some gauge groups not mentioned in \rChP.
We also explain why the set of possible gauge groups
in four dimensions turns out to be self-dual.

\subsec{Gauge Groups and Roots.}

As we have discussed in the previous section,
the moduli space of $9-d$-dimensional compactification
is locally the Grassmanian  of $d+1$-dimensional spaces
in $\IR^{d+1,d+9}$. Given such a plane $\nu$,
one can write a mass formula:
\eqn\massproj{\matrix{
M^2(p,N',\tilde{N})=
{1\over 4}({\cal P}_{\perp} p)^2-{1\over 4}({\cal P}_{\|} p)^2+
N'+\tilde{N}-1=\cr
=\half ({\cal P}_{\perp}p)^2+2N'-2=-\half ({\cal P}_{\|}p)^2+\tilde{N}
}}
for the state with the momentum $p=(\{l_j\},\{n_j\},R)$,
where the last two equalities follow from the level matching
condition. 
We have denoted  ${\cal P}_{\|}$ and ${\cal P}_{\perp}$
the projection on the plane and the orthogonal projection,
respectively. It follows that the state is massless if and only if
\eqn\massless{\matrix{
{\cal P}_{\|}p=0,\cr \tilde{N}=0,\cr
{1\over 4} p^2 +N'=1
}}
The first condition means that the plane
 should be orthogonal to the momentum of the massless state.
Notice that for the state of the class {\bf 1}, the minimal
allowed value of $N'$ is zero (the expansion of \first\ 
in powers of $q$ starts with $q^0$), thus the states with $p$ of
class {\bf 1} and $p^2=4$ will be massless\foot{
In the lattice $\gii\oplus\ge$, corresponding to the nine-dimensional
theory, the length square of each level two
vector is divisible by eight, but
in lower dimensions we may have vectors of class {\bf 1} with
length square four, because of the factor $\Gamma_{d,d}(2)$.}.
For the vectors of class {\bf 2}, the minimal value of $N'$
is $1$, and we the corresponding state cannot be massless.
For the vectors of class {\bf 3}, the minimal value
of $N'$ is $\half$, and it occurs with multiplicity one, as follows
from \third\  (or \second). Thus, we get one massless state
for each $p$ with $p^2=2$. 

Thus, the massless states correspond to either vectors of
length square $2$ (such vectors can actually have only level one),
or level two vectors with length square $4$. This is very natural,
since precisely to these two types of vectors we can associate
a reflection: it is given by the formula
$$
r_v(w)=w-2{(v\cdot w)\over (v\cdot v)}v
$$
--- this transformation is a symmetry of the lattice if and only if
half of the length square of the vector $v$ divides its level.

 If at some point of the moduli space such 
a state becomes massless, we get an enhanced gauge symmetry.
The corresponding plane in $Gr(d+1,\IR^{d+1,d+9})$ is invariant under the 
reflection in the momentum vector of the massless state. 
Those points in the moduli space which correspond to the
non-invariant planes may be obtained by perturbing the
action by the current in the Cartan subalgebra of the enhanced $SU(2)$.
The T-duality corresponds to the Weyl group of the enhanced gauge group
\rGPR. 

We have to stress the essential difference with the description of the
gauge groups for the usual heterotic compactification. Given the background
specified by the plane $\omega\in\IR\times\Gamma_{(D)}$, the roots of the
gauge group belong to the sublattice $\omega^{\perp}\subset\Gamma_{(D)}$.
For the usual heterotic string, the root system consisted of all the vectors
of length square 2 in this sublattice. For the CHL compactifications,
the root is either the vector with the length square 2 or the vector
with the length square 4 which is on the level two {\it in the whole 
lattice}. Thus, given just the lattice $\omega^{\perp}$, one cannot yet
tell what is the gauge group: one has to know the embedding 
$\omega^{\perp}\in\Gamma_{(D)}$. 

It was shown in Section 3.2, that vectors of length square 2 correspond
to either states from the twisted sector, or states from the untwisted
sector with $P-Q\in\Delta+2\Gamma_8$. The massless states from the level
two sublattice correspond to states from untwisted sector with $P=Q$.
Comparing this to the description of the world-sheet Kac-Moody algebras
in section 2, we see that those massless states whose
momentum has length square 2 correspond to the level 2 $su(2)$ Kac-Moody
algebra, while the length square 4 vectors correspond to the
level 1 Kac-Moody algebra.

\subsec{Examples of Gauge Groups.}

Before proceeding with the gauge groups, let us remind here
the definitions of some root systems. Unfortunately, the root
systems are usually denoted by the same letters as the root lattices.
To avoid confusion, we will use the bold letters for the
root systems, and the usual letters for the root lattices (the lattices,
generated by roots).  
Consider the space 
$\IR^n$ with the basis $\{u_i\}_{i=1,\ldots,n}$, 
$(u_i\cdot u_j)=\delta_{ij}$. We will need the following root systems:

1) $\Br_n$: consists of the vectors of the form $\stw(\pm u_i\pm' u_j)$,
and the vectors $\pm \stw u_i$. Corresponds to the algebra $so(2n+1)$.
The root lattice is $B_n\simeq \IZ^n(2)$. Notice that we use the
normalization of the roots of $\Br_n$, which differs from the
normalization usually accepted in the textbooks on Lie groups \rVinberg\
by the factor of $\stw$. Our normalization agrees with the embedding
of the root system in the CHL momentum lattice.

2) $\Cr_n$: vectors of the form $\pm u_i\pm' u_j$, plus $\pm 2u_i$.
Roots of the Lie algebra $sp(2n)$ (sometimes called $sp(n)$).
They generate the root lattice $C_n\simeq D_n$.

3) $\Dr_n$: $\pm u_i\pm' u_j$, $so(2n)$, the root lattice $D_n$.

4) $\Fr_4$: roots of $C_4$ plus $\epsilon_1 u_1+\epsilon_2 u_2+
\epsilon_3 u_3+\epsilon_4 u_4$ with all the possible signs 
$\epsilon_1,\ldots,\epsilon_4$. Root lattice $F_4\simeq D_4$.

We have chosen such a normalization of the roots that the long
roots have length square 4. As was explained at the end of the
last subsection, this corresponds to the level one Kac-Moody
algebra. 

In eight dimensions we can embed the root system
$\Dr_9$ in the lattice $\gii\oplus \ge$. Indeed, consider the
root system for this lattice (\rConway, Ch. 27). The corresponding
Dynkin diagram may be obtained from the affine diagram for $E_8$ 
by attaching one more node to the affine root.

\kartinka{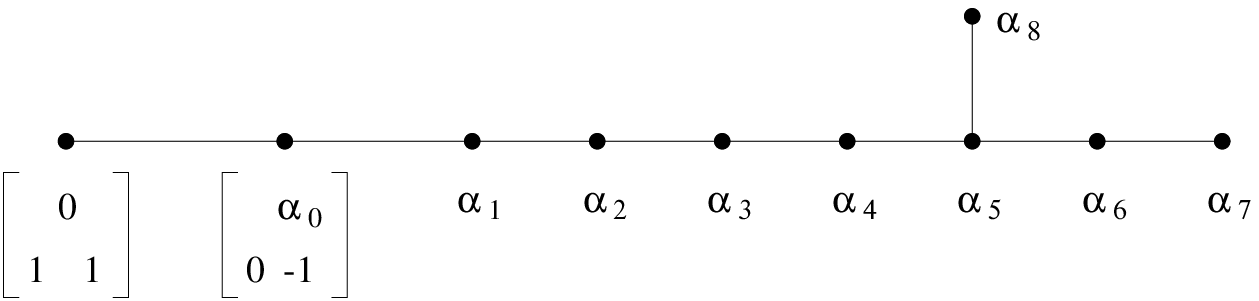}{Root system for $II_{1,9}$.}{2}

If we remove the seventh root of $E_8$, we
get the Dynkin diagram for the root system $\Dr_9$. Thus, at the 
corresponding point of the moduli space we get the gauge algebra $so(18)$.

In eight dimensions, on can get the algebra $sp(20)$ in the following
way. Represent the lattice as $\Gamma_{2,2}\oplus D_8$.
Take the $D_8$ sublattice whose orthogonal complement is $\Gamma_{2,2}$.
The root system $\Cr_8$ can be embedded in the lattice $D_8$, so
that its long roots are on the level two (indeed, the lattice
generated by the root system $\Cr_n$ is $D_n$).
To get a root system $\Cr_{10}$, we need just to add two vectors
from $\Gamma_{2,2}$:

\risunok{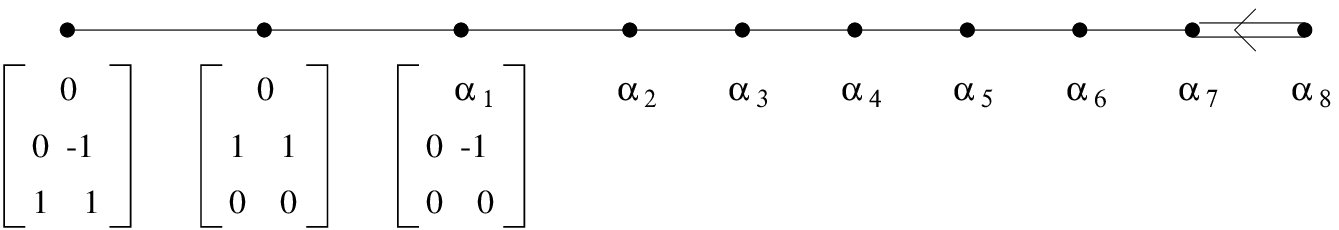}{Embedding the root system of $sp(20)$.}{3}

In the case $d\geq 2$ we can actually embed the root system of
$sp(20)_1\oplus so(2d-1)_1$. 
Indeed, let us represent the lattice in the following form:
\eqn\forSpSO{
\Gamma_{d-1,d-1}(2)\oplus\Gamma_{2,2}\oplus D_8
}

First, let us embed $\Cr_{10}$ in $\Gamma_{2,2}\oplus D_8$,
as described above. To find the root system  $\Br_{d-1}$,
let us consider the root system ${\rm\bf A}_{d-1}(2)$
(the roots of $su(d)$, rescaled by $\sqrt{2}$), embedded into
$\Gamma_{d-1,d-1}(2)$ \rNarain. 
We can take the first $d-2$ vectors of this root
system to be the long roots of $\Br_{d-1}$, and as a short root we
may choose the last root of ${\rm\bf A}_{d-1}(2)$ plus the vector
$$
\vtrd{0}{0}{\;\;0}{1}{-1}\in \Gamma_{2,2}\oplus D_8
$$
which is orthogonal to all the roots of $sp(20)$.  
This gives the root system of $sp(20)_1\oplus so(2d-1)_1$.

We can also get the gauge group $sp(20-2n)\oplus so(2d+2n-1)$
for any $n\leq d+1$. To find these gauge groups, we will need
another useful isomorphism\foot{The way we get the roots of
$sp(20-2n)\oplus so(2d+2n-1)$ is essentially the translation
to our language of the method suggested in \rChP.}. 
First of all, it is well known
\rNarain, that $\Gamma_{n,n}$ is isomorphic to the lattice 
consisting of the pairs $(w;\tilde{w})$, such that $w,\tilde{w}\in D_n^*$ 
and $\tilde{w}-w\in D_n$. It turns out, that the lattice 
$\Gamma_{n-1,n-1}(2)\oplus\gii$ is isomorphic to the lattice, 
consisting of the vectors $\stw(w;\tilde{w})$, where 
$w,\tilde{w}\in D_n^*$, and the difference
$w-\tilde{w}$ is either in $D_n$ or 
in the vector conjugacy class of $D^*_n$ modulo $D_n$.  This is 
explained in Appendix A. 
Notice an embedding
\eqn\Bn{
\IZ^n(2)\oplus \IZ^n(-2)\subset \Gamma_{n-1,n-1}(2)\oplus\Gamma_{1,1}
}
This embedding has the property that the long roots of both
$B_n$ and $B_n(-1)$ are on the level two in the whole lattice
$\Gamma_{n-1,n-1}(2)\oplus\gii$. 

Then, for $n>0$ consider the embedding
\eqn\g{
\IZ^{n-1}(4)\oplus D_{10-n}\subset (\gii\oplus D_8)
\subset (\gii\oplus D_8)\oplus\Gamma_{d-1,d-1}(2)\oplus\gii
}
where all the vectors in $\IZ^{n-1}(4)$ are on the level two
in $\Gamma_{(D)}$.
In the lattice $\IZ^{n-1}(4)$, let us pick $n-1$ vectors $f_j$ 
with the scalar products
$$
f_i\cdot f_j=4\delta_{ij}
$$ 
Using the embedding \Bn, we can find in
$\Gamma_{d-1,d-1}(2)\oplus\gii$ the $2d$ vectors 
$e^+_j$ and $e^-_j$ with the scalar products
$$
e^+_i\cdot e^+_j=2\delta_{ij},\;\; e^+_i\cdot e^-_j=0,\;\;
e^-_i\cdot e^-_j=-2\delta_{ij}
$$
Then, the vectors $e^+_j$ for $j=1,\ldots, d$ and 
$f_j+e^-_j$ for $j=1,\ldots, n-1$ may be taken as the short
roots of $so(2d+2n-1)$ on the level one. The long roots are
expressed as sums and differences of the short roots. The root system
of $sp(20-2n)$ is embedded into the lattice $D_{10-n}$ on the left hand
side of \g.

We can find gauge groups $\Fr_4$ in the seven-dimensional theory. 
Indeed, let us use the isomorphism 
$$
\Gamma_{(7)}\simeq \Gamma_{3,3}\oplus D_4\oplus D_4
$$
It follows that in the seven-dimensional compactification
we can get the gauge algebra
\eqn\seven{
so(6)_2\oplus (\Fr_4)_1\oplus (\Fr_4)_1
}
Indeed, we can embed the root system  $\Dr_3$ into $\Gamma_{3,3}$
by embedding it into the lattice $D_3$, which is 
a sublattice of $\Gamma_{3,3}$, as we explained above. 
Notice that, although the lattice
$D_3\subset \Gamma_{3,3}$ contains the root system $\Cr_3\supset \Dr_3$,
we do not get the algebra $sp(6)$, but only $so(6)$. The reason is that
the long roots of $sp(6)$ are not on the level two in the whole
lattice $\Gamma_{3,3}$. The root system $\Fr_4$ is embedded into 
the lattice $D_4$.

For the six-dimensional compactification, we can embed 
\eqn\six{
sp(8)_1\oplus (\Fr_4)_1\oplus (\Fr_4)_1
}
Indeed, the simple roots of the root system $\Cr_4$ of $sp(8)$ fit into 
$\Gamma_{3,3}\oplus\gii(2)$ in the following way:
\risunok{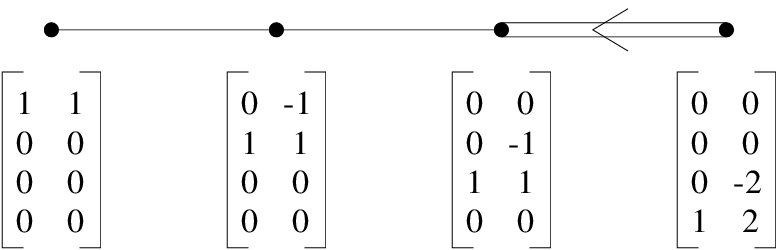}{Embedding $\Cr_4$ into 
$\Gamma_{3,3}\oplus\Gamma_{1,1}(2)$.}{4}
Here the first three rows of the matrix are the coordinates in 
$\Gamma_{3,3}$, and the last one is the coordinates in 
$\Gamma_{1,1}(2)$. One can see that the long roots are on the level two.

It was argued in \rChP, that the gauge algebra 
$so(9)\oplus \Fr_4\oplus\Fr_4$ appears in the six-dimensional 
compactification. This is in disagreement with our results:
it is impossible
to embed the root system $\Br_4\oplus\Fr_4\oplus\Fr_4$ into our lattice
$\Gamma_{(6)}$.

\subsec{Topology of Gauge Groups.}

We should comment on the global structure of these gauge groups.
Let us first remind a few basic facts from the theory of Lie
groups \rVinberg.
Suppose ${\bf h}$ is a Cartan subalgebra of the semisimple
Lie algebra ${\bf g}$. Since any two elements from ${\bf h}$
commute, the corresponding adjoint operators may be simultaneously
diagonalized. The eigenvalues are the linear functionals on ${\bf h}$,
that is the elements of ${\bf h}^*$. They are called roots.
For each root $\alpha_i$, we have the corresponding $sl(2)$ subalgebra
$\{e_{\alpha_i},f_{\alpha_i},h_{\alpha_i}=
\left[ e_{\alpha_i},f_{\alpha_i} \right] \}$. 
Elements $\alpha_i^{\vee}=\kappa_i h_{\alpha_i}\in{\bf h}$, where
$\kappa$ is the coefficient adjusted in such a way that
$\left[\alpha_i^{\vee},e_i\right]=2e_i$, are called {\it coroots}.
Notice
\eqn\normCoroot{
\alpha_i(\alpha^{\vee}_i)=2
}
There is an invariant bilinear form on ${\bf g}$, which enables
us to identify ${\bf h}$ with ${\bf h}^*$. We will call
the corresponding isomorphism of linear spaces 
$\nu: {\bf h}^*\tilde{\to}{\bf h}$.
We have:
\eqn\RootCoroot{
\alpha_i^{\vee}={2\over (\alpha_i,\alpha_i)}\nu(\alpha_i)
}
Let us explain this formula. Consider $h\in{\bf h}$ such that
$\left[ h,e_{\alpha_i}\right] =0$. This means that $h\perp\nu(\alpha)$.
Let us prove $h\perp\alpha^{\vee}$. Indeed, 
$$(h\cdot\alpha_i^{\vee})=\kappa^{-1}(h\cdot\left[ e_{\alpha_i},
f_{\alpha_i} \right])=\kappa^{-1}(\left[ h,e_{\alpha_i}\right]
\cdot f_{\alpha_i}) =0$$
Since an arbitrary element $h\in{\bf h}$ orthogonal to $\nu(\alpha)$
is orthogonal also to $\alpha^{\vee}$, we get 
$\nu(\alpha)\sim\alpha^{\vee}$. The coefficient of proportionality 
may be found from \normCoroot.

Let us denote $Q$ the lattice generated by the roots, and $Q^{\vee}$
the lattice generated by the coroots. The weight lattice 
$$P=(Q^{\vee})^*$$
consists of those vectors which have integer scalar products with
the coroots. 

The global structure of the Lie group $G$ is defined by the lattice
 $$X(T)={\rm Hom}(T,S^1)$$
--- the lattice of characters of the maximal torus. This may be
an arbitrary sublattice of $P$, containing $Q$:
\eqn\Xbetween{
Q\subset X(T)\subset P
}

The first homotopy group $\pi_1(G)$ of $G$ is isomorphic
to $P/X(T)$.

Given the Lie algebra $\bf g$ with the root system $\Delta$,
 we can define the dual Lie algebra ${\bf g}^{\vee}$ with the
dual root system $\Delta^{\vee}$. The topology of the dual Lie group
is specified as follows: $X(T_{G^{\vee}})=(X(T_G))^*$.
Examples of pairs $(G,G^{\vee})$ may be found in \rGNO.

If at some point of the moduli space we have a root system 
$\Delta$ embedded into the sublattice orthogonal to $\omega$, 
then we get the corresponding group $G$ as the gauge group.
Some information about the global structure of the gauge
group may be obtained from the known structure of the
perturbative spectrum. Consider the orthogonal projection
$\pi$ on the space, generated by $\Delta$. The perturbative
states are charged under the group $G$. The charges span
the lattice $\pi \Gamma_{(D)}$. This means, that at least
\eqn\Xsupset{
X(T)\supset \pi\Gamma_{(D)}
}
which implies
$$
\pi_1(G)\subset P/\pi\Gamma_{(D)}
$$

{\bf Examples.} We have found the embeddings into $\Gamma_{(D)}$ of the root
systems $B_n$ for various $n$. The
coroot lattice of $B_n$ is $B_n^{\vee}=C_n\left(\half\right)$. 
$C_n\left(\half\right)^*$ is generated
by the vectors $\stw u_j$, $j=1,\ldots,n$, and the ``spinor''
$\stwi\sum_j u_j$. One can see that the ``spinor'' weight does
actually belong to $\pi \Gamma_{(D)}$. Indeed, the spinor weight 
of $B_n$ is characterized by the property that its scalar product
with the short roots of $B_n$ is 1. Since the short roots
of $B_n$ are on the level one in $\Gamma_{(D)}$, there is at
least one vector $v\in\Gamma_{(D)}$ such that the scalar product 
of $v$ with the given short root is one. The projection $\pi v$
of this vector belongs to the spinor conjugacy class. 
This means that the 
group is in fact $Spin(2n+1)$ --- the universal covering of
$SO(2n+1)$. 

The root system $sp(20)$, embedded into the eight-dimensional lattice
as shown on Fig.3, corresponds to the group $Sp(20)$ (the simply
connected group). Indeed, consider the vector $\vtrd{0}{0}{0}{0}{1}$.
The scalar product of this vector with all the roots of $sp(20)$
is zero, except for the scalar product with the leftmost root,
which is equal to one. This means, that the corresponding state
transforms in the fundamental representation of $Sp(20)$.
On the other hand, for the root system $sp(20-2n)$  for
$n>1$, embedded in the sublattice $D_{10-n}$ in \g, there
are no states in the perturbative spectrum, transforming in
the fundamental representation. This suggests that the corresponding group
is actually $Sp(20-2n)/\IZ_2$. 

\subsec{S-duality in four dimensions.}

It was observed in \rChP, that the set of maximal gauge groups in 
four dimensions is
self-dual: if we have the gauge group $G$ at some point
in the moduli space, we must have the gauge group $G^{\vee}$
at some other point in the moduli space. 
In this subsection we will give an explanation of this fact, 
using our description of the lattice.

There is an involution on the
moduli space, which transforms the background with
the gauge symmetry $G$ to the background with the gauge symmetry $G^{\vee}$.
This involution may be interpreted as the action of S duality on the
background fields.
To explain how S duality works, we have to remember that string
theory has infinitely many nonperturbative states, arising from
the quantization of the solitons. In this paper we have so far been
concerned with the perturbative states only.  In the low energy
theory, we have $20$ abelian gauge fields at the general point in the
moduli space. Perturbative states arising from strings having nonzero
momenta or winding in the compactified directions are electrically
charged under these gauge fields. From this point of view $\Gamma_{(4)}$
is the lattice of electric charges. There are also nonperturbative states,
carrying magnetic charges. As explained in \rSchwarzSenHet, they correspond
to fivebranes partially wrapping the six-torus.  Given the lattice of
electric charges, the allowed magnetic charges should satisfy 
certain quantization conditions (see \rSchwarzSenHet\ and references
therein). These conditions require that the allowed magnetic charges
belong to the lattice $\Gamma_{(4)}^*$. 

It was conjectured in \rChP, that the given
four-dimensional CHL theory with the coupling constant $\lambda$ 
is equivalent to the dual CHL theory with the coupling constant 
$1\over\lambda$,  so that the perturbative states of the dual theory are
identified with the magnetically charged nonperturbative
states in the original theory.  The background fields are specified
by the plane $\omega\subset\IR\otimes\Gamma_{(4)}$. From the point
of view of the dual theory, $\omega$ is the plane in 
$\IR\otimes\Gamma_{(4)}^*$. An important property of the four-dimensional
CHL lattice is that $\Gamma_{(4)}\simeq\Gamma_{(4)}^*(2)$. 
Indeed,
\eqn\selfD{
\Gamma_{(4)}\simeq \Gamma_{3,3}\oplus\Gamma_{3,3}(2)\oplus D_4\oplus D_4
\simeq \Gamma_{(4)}^*(2)
}
We have $\Gamma_{3,3}(2)^*\simeq\Gamma_{3,3}\left(\half\right)$ since
$\Gamma_{n,n}$ is self-dual, and
the isomorphism $D_4^*\simeq D_4\left(\half\right)$ is 
explained\foot{Let us explain it here for completeness. For the lattice 
$D_4=\{n_1u_1+\cdots+n_4u_4|n_1+\cdots+n_4\in 2\IZ\}$, the
weight lattice $D_4^*$ consists of the vectors $m_1u_1+\cdots m_4u_4$,
where all $m_j$ are either simultaneously integers, or simultaneously
half-integers. The simple roots $\alpha_0,\ldots,\alpha_4$ 
of $\widehat{D}_4$  (satisfying the relation 
$\alpha_0+2\alpha_1+\alpha_2+\alpha_3+\alpha_4=0$) are embedded into 
$D_4^*(2)$ in the following way: $\alpha_0=\stwi(u_1-u_2-u_3-u_4)$,
$\alpha_1=\stwi(u_1+u_2+u_3+u_4)$, 
$\alpha_2=\stwi(u_2-u_1-u_3-u_4)$, $\alpha_3=\stwi(u_3-u_1-u_2-u_4)$,
$\alpha_4=\stwi(u_4-u_1-u_2-u_3)$; they generate the lattice $D_4^*(2)$.}
on pp.118-119 of \rConway. Let us fix an isomorphism 
$S: \Gamma_{(4)}^*\to\Gamma_{(4)}\left(\half\right)$ (they
are all conjugate by symmetries of $\Gamma_{(4)}$, which
will give equivalent backgrounds). Consider the 
background specified by the plane $\omega\subset\IR\otimes\Gamma_{(4)}$. 
Since $\Gamma_{(4)}\subset\Gamma_{(4)}^*$, we may consider this
embedding as an embedding of $\omega$ into $\IR\otimes\Gamma_{(4)}^*$.
Acting on $\omega$ by the operator $S\in O(6,14)$, we get
\eqn\Iomega{
\omega\subset\IR\otimes\Gamma_{(4)}\left(\half\right)
}
which is the same as an embedding of $\omega$ into $\IR\otimes\Gamma_{(4)}$.
This gives the required transformation of the background. 
This is an involution ($S^2\omega=\omega$), commuting with the T dualities.

Suppose that we start with the background, specified by the
plane $\omega$, such that the sublattice $\omega^{\perp}\cap \Gamma_{(4)}$ 
contains a root system $\Delta$ (the long roots of this root system should
be on level two in the whole lattice $\Gamma_{(4)}$, not only
in $\omega^{\perp}\cap \Gamma_{(4)}$). We want to prove that
the lattice  $\omega^{\perp}\cap\Gamma_{(4)}^*$ contains
the dual root system $\Delta^{\vee}$. 
To obtain the dual root system $\Delta^{\vee}$, we have
to divide by 2 the long roots of $\Delta$, and make the
overall rescaling (this rescaling is related to the
coefficient $\half$ in \Iomega).  The long roots of $\Delta$ are on
the level 2 in the whole lattice $\Gamma_{(4)}$. This implies
that half of a long root belongs to $\Gamma_{(4)}^*$. 
The short roots of $\Delta$ have integer scalar products
with all the vectors from $\Gamma_{(4)}$; after rescaling
of the metric by 2 they become the level two vectors in 
$\Gamma_{(4)}^*(2)=\Gamma_{(4)}$. Thus, they may be taken as
the long roots of $\Delta^{\vee}$. This proves that 
$\Delta^{\vee}\subset (S\omega)^{\perp}$.

If the root system $\Delta$ was maximal in $\omega^{\perp}$
(that is, we could not find any other root system, containing $\Delta$
as a proper subset), then the root system $\Delta^{\vee}$ in
$\omega^{\perp}\cap \Gamma_{(4)}$ is also maximal. Indeed,
suppose that there is some larger root system 
$\hat{\Delta}\supset\Delta^{\vee}$ in $\omega^{\perp}\cap \Gamma_{(4)}^*$.
Then, we may use $S^2=1$ to prove that $\omega^{\perp}\cap\Gamma_{(4)}$
contains $\hat{\Delta}^{\vee}$. But $\hat{\Delta}^{\vee}$ is
larger then $\Delta$, which contradicts to the assumption that
$\Delta$ is maximal.

\newsec{Comparison to Type IIA, M Theory and F Theory Duals.}

\subsec{Type IIA.}

It was argued by Schwarz and Sen \rSchwarzSen, that the six-dimensional 
compactification of the CHL string is dual to Type IIA
on the singular K3 surface $Y$, with some Ramond-Ramond background fields
turned on. The K3 surface involved is the quotient of a smooth
K3 $X$ by certain $\IZ_2$ involution. The corresponding involution of
the cohomology lattice $H^2(X,\IZ)=II_{3,19}$ is the exchange
of the two $\Gamma_8$ sublattices. Also, the RR background field
should be turned on. The origin of this RR background may be explained
from the point of view of M Theory lift. Type IIA on K3 is M Theory
on ${\rm K3}\times S^1$. The theory of Schwarz and Sen
is the quotient by the following symmetry \rSchwarzSen: shift by half of
the M Theory circle, and exchange of two $\Gamma_8$ sublattices
in the cohomology group of K3.
In this section we study the moduli space of Type IIA on
K3 with involution, and show that it coincides with our 
answer for the moduli space of six-dimensional CHL string.
This is an evidence in favour of the proposed duality.

We will consider the $\IZ_2$ involution $\iota$  of the smooth
K3 surface $X$, which acts as an identity
on $H^{2,0}(X)$. Such involutions are known in algebraic geometry 
as Nikulin involutions \refs{\rNikulinKt,\rMorrison}; we give an
example of the Nikulin involution in Appendix B. 
More precisely,
consider the symmetry $\iota^*$ 
of the lattice $L=H^2(X,\IZ)\simeq \Gamma_{3,19}$,
exchanging the two $\ge$ sublattices in $\Gamma_{3,19}\simeq
\Gamma_{3,3}\oplus \ge\oplus\ge$. The moduli space of Type IIA
worldsheet conformal field theory is parametrized by the four-planes
in $\IR\otimes II_{4,20}$. If the plane is orthogonal to
the antidiagonal $(\ge)_{Anti-Diag}\subset \ge\oplus\ge\subset II_{4,20}$,
then
the symmetry $\iota^*$ is related to the symmetry of the worldsheet 
conformal field theory. 
Indeed, the complex structure of $X$
specifies an oriented 2-plane $\omega$ in $\IR\otimes H_2(X,\IZ)$
(through the period map). This plane is a subspace of the 4-plane
which specifies the background.
Suppose that this plane is orthogonal to the anti-diagonal 
$(\ge)_{Anti-Diag}\subset \ge\oplus\ge\subset L$. This means that
$\iota^*$ preserves
the Hodge structure of $\;\;\IC\otimes H_2(X,\IZ)$, 
specified by the plane $\omega$.
It also preserves the cones $V^+$ and $V^-$ defined as 
$V^+\cup V^-=\{x\in\IR\otimes H_2(X,\IZ)|x^2>0\}$, and it leaves all
the vectors with length square $-2$ invariant. Under these conditions,
the Theorem 2.7' from \rNikulinKt\ tells us that there exists an algebraic 
automorphism $\iota$ of $X$ (an involution), whose action in cohomology 
coincides with $\iota^*$. Since the Kahler classes of all the cycles
in $H^2(K3,\IZ)$ are invariant under $\iota^*$, the metric
is invariant under $\iota$ (since there
is only one metric with the given Kahler class). The way our 4-plane
parametrizes the $B$ field and the volume of K3 is explained in \rAspinwall.
It follows that the $B$-field is $\iota$-invariant, if the plane
is diagonal.

This means that if the point of the moduli space
$O(II_{4,20})\backslash O(4,20)/O(4)\times O(20)$ is represented
by the plane orthogonal to $(\ge)_{Anti-Diag}\subset 
\Gamma_{4,4}\oplus \ge\oplus\ge$, then the symmetry of the
lattice exchanging the two $\ge$ sublattices is a symmetry
of the worldsheet theory, and we may consider the corresponding 
orbifold. Considering just an orbifold $K3/\iota$ does not
give a new theory, because $K3/\iota$ is birational to another
$K3$: what we would get is again $IIA/K3$.  To get the Type IIA
dual of CHL string we have, as prescribed in \rSchwarzSen, to turn
on the RR flux localized on the fixed points of $\iota$.
The quotient $X/\iota$ is
singular, because $\iota$ has eight fixed points. 
To study the geometry of the quotient, it is convenient first to blow up
these eight singular points, and get a smooth $K3$ surface $Y$.
Then the minimal primitive sublattice
of $H^2(Y,\IZ)$ containing the exceptional curves on $Y$ would be isomorphic
to $D_8^*(-2)$ \refs{\rNikulinKt,\rMorrison}. 
This lattice is called Nikulin lattice, and we introduce for
convenience a notation: $$\nik=D_8^*(-2)$$
In Appendix B, we explain, in one particular case, why the singularities
generate this lattice.

The RR background field may be defined in the following way.
Let $Y'$ be $Y$ with the eight exceptional divisors thrown away.
This manifold has nontrivial first homology group \rNikulinKt:
$$
H_1(Y',\IZ)=\IZ_2
$$
Then, the monodromy of the RR gauge field around the
generator of $H_1(Y',\IZ)$ should be $-1$. It was argued
in \rSchwarzSen, that turning on this background makes
the singularity irremovable.

The moduli space for $K3$ with irremovable singularities,
whose vanishing cycles generate the Nikulin lattice,
is parametrized by those planes
in $Gr(4,\IR^{4,20})$ which are orthogonal to the Nikulin lattice. 
The orthogonality condition ensures that the vanishing cycles
are collapsed, and there is no $B$-field flow through them. 
Thus, the moduli space of our theory 
is locally just the locus in the moduli space of IIA$/K3$, 
corresponding to the planes orthogonal to the Nikulin lattice. 
Which discrete identifications should we make? 
We have to consider those symmetries of the lattice 
$II_{4,20}$, which preserve the sublattice generated by
singularities. There
is a unique primitive embedding of the Nikulin lattice 
into $II_{4,20}$ (this follows from the Theorem 2.8 in
\rMorrison, we put this theorem in Appendix A for convenience), 
and the orthogonal sublattice is 
$\Gamma_{(6)}=\Gamma_{4,4}\oplus D_8^*(-2)$. The discrete symmetries
of the background are the symmetries of this lattice.
Indeed, any symmetry of the lattice $II_{4,20}$ which preserves
the lattice generated by singularities acts correctly
on the orthogonal lattice $\Gamma_{(6)}$. The converse
is also true: given the symmetry of the sublattice $\Gamma_{(6)}$,
we may continue it to the symmetry of $II_{4,20}$.
The last statement is, in fact, not completely trivial,
since it is not true 
that $II_{4,20}$ is the direct sum of the Nikulin lattice
and the lattice $\Gamma_{(6)}=\Gamma_{4,4}\oplus D_8^*(-2)$
(such a direct sum would not be a self-dual lattice). 
This means, that generically speaking, we cannot just
continue the symmetry of $\Gamma_{(6)}$ as identity 
on the orthogonal sublattice: this would not preserve
the way $\Gamma_{(6)}$ and $\Gamma_{(6)}^{\perp}$ are ``glued
together'' in $II_{4,20}$  (in other words, such a naive
continuation would not be a symmetry of the lattice.)
Let us explain how to find a correct continuation. To get
a self-dual lattice $L$ from the sum of two non-self-dual lattices
$M_1$ and $M_2$, we have to take generators of the form 
$px+qy$ where  $x\in M_1$, $y\in M_2$, but $p,q$ are some
{\it rational} numbers. The vectors in the self-dual lattice 
modulo those in the direct sum $M_1\oplus M_2$ are called 
{\it the glue vectors}. For these vectors $px\in M_1^*$,
and $qx\in M_2^*$.
In fact, one can prove \rConway,
that the map $\gamma: \left[px\right]\to\left[qy\right]$ 
from $M_1^*/M_1$ to $M_2^*/M_2$ is an isomorphism. 
We want to continue the symmetry $g_1\in O(\Gamma_{(6)})$
to the symmetry of the lattice $II_{4,20}$. Most
of the symmetries of $\Gamma_{(6)}$ do not
act as an identity on $A_{\gsix}=\gsix^*/\gsix$. Thus, we
cannot continue  such a symmetry $g_1$
just as an identity on $(\gsix)^{\perp}=\nik$:
what we should do instead is to find such a symmetry 
$g_2 \in O(\nik)$ which reduces on $\nik^*/\nik$ to
$\gamma g_1 \gamma^{-1}$, and the proper continuation of
$g_1$ will be then $g_1\oplus g_2$. In fact, we can always
find such a $g_2$, because as we prove in the Appendix A,
an arbitrary symmetry $\bar{g}$ of $\nik^*/\nik$ may be
represented by the symmetry $g$ of $\nik$.

This gives the moduli space
isomorphic to the Grassmanian modulo $O(\Gamma_{(6)})$,
which coincides with our result for the
momentum lattice for the six-dimensional compactification of the
CHL string. 

Notice that the discrete identifications corresponding to 
{\it geometric} symmetries of $Y$ (those which do not involve
the $B$ field) may be explained in somewhat simpler way. 
Indeed, the geometric moduli space of $Y$ with Nikulin
singularities are the same as the moduli space of those
smooth K3 surfaces $X$, which admit Nikulin involution. 
This condition means that the corresponding plane
in $\IR^{3,19}$ should be diagonal. The subgroup
of $\Aut(II_{3,19})$ which preserves the anti-diagonal
$E_8$ is $\Aut(\Gamma_{3,3}\oplus \ge(2))$ (we have to prove
that any symmetry of the sublattice 
$\Gamma_{3,3}\oplus\ge(2)\subset II_{3,19}$ may be extended
to the symmetry of $II_{3,19}$; this may be derived from
the surjectivity of the map $\Aut(\ge(2))\to \Aut(A_{\ge(2)})$,
which is proven in the Appendix A.) It follows from
\Auts\ and $\Gamma_{3,3}(2)\oplus\ge\simeq \Gamma_{3,3}\oplus\nik$,
that the group of geometric symmetries coincides with
$$
\Aut(\Gamma_{3,3}\oplus\nik)
$$
which is in agreement with what we have obtained from the
study of the theory on the quotient surface $Y$. This may be considered
as a geometric interpretation of the isomorphism 
$\Gamma_{3,3}(2)\oplus\ge\simeq\Gamma_{3,3}\oplus{\cal N}$.

\subsec{M Theory and F Theory Duals.}
Recently, it was proposed \refs{\rLL,\rWittenZt}, 
that the seven-dimensional
CHL string is dual to the compactification of M theory
on K3 surface with some irremovable $D_4\oplus D_4$ singularity,
and  the eight-dimensional compactification may be described
as F theory on K3 with irremovable $D_8$ singularity \rWittenZt. 
Let us show,
that this is in agreement with our description of the moduli space
of CHL string.

Consider first $D=8$. The moduli of the F-theory 
compactification are parametrized by the complex structures of
the elliptic K3 surfaces with a section \rVafa, which may be thought of as 
the timelike planes in $\IR^{2,18}$, modulo the action of the
discrete group of symmetries $II_{2,18}$. 
The $D_8$ singularity means that
the rational cycles, with the intersection matrix realizing
the root system of $D_8$, become holomorphic. This means, that 
the plane in $\IR^{2,18}$ becomes orthogonal to the roots of 
$D_8$. The primitive embedding of the root system of $D_8$
into $II_{2,18}$ is unique modulo the symmetries of $II_{2,18}$ 
(see Theorem A1 in the Appendix A.). It may be constructed 
in the following way: 
\eqn\embei{
D_8\subset\Gamma_{16}\subset\Gamma_{2,2}\oplus\Gamma_{16}
\simeq II_{2,18}
}
From this representation of the embedding, one can see that the 
orthogonal complement to this $D_8$ is 
$$
D_8^{\perp}=\Gamma_{2,2}\oplus D_8
$$
which is our result for the Narain lattice of the eight-dimensional
CHL theory. One can prove, that an arbitrary isomorphism of 
$D_8^{\perp}\subset II_{2,18}$ may be extended 
to the isomorphism of $II_{2,18}$. This means, that the discrete
symmetries of the F Theory background are the symmetries of 
the lattice $D_8^{\perp}=\Gamma_{(8)}$. Thus, the moduli space
of the CHL string in eight dimensions is isomorphic to the
moduli space of its F Theory dual.

In the recent papers \refs{\rBPS,\rBKMT}, 
it was argued that the moduli space
of the eight-dimensional CHL theory is isomorphic to the moduli
space of Eriques surfaces, which is the arithmetic quotient:
\eqn\Enriques{
O(\Gamma)\backslash O(2,10)/O(2)\times O(10)
}
where $\Gamma=\gii\oplus\gii(2)\oplus\ge(2)$. It follows from the
isomorphism $\Gamma^*\simeq \Gamma_{(8)}\left(\half\right)$
and the fact that $O(L)\simeq O(L^*)$ for an arbitrary lattice 
$L$, that the answer \Enriques\ for the moduli space
coincides with our answer \Mcal\ for the eight-dimensional
theory ($d=1$).

Consider $D=7$. The dual theory is M Theory compactified on
$K3$ with irremovable $D_4\oplus D_4$ singularity.
It follows from the Theorem A1 in the
Appendix A, that there is a unique primitive embedding
of $D_4\oplus D_4$ into $II_{3,19}$.
We can represent it as follows:
\eqn\embse{
D_4\oplus D_4\subset E_8\oplus E_8\subset E_8\oplus E_8
\oplus \Gamma_{3,3}\simeq II_{3,19}
}
Its orthogonal complement 
is:\foot{Notice that in all three cases, embedding the 
lattice $X$ generated by singularities into the unimodular
lattice, we have got the orthogonal complement to $X$ isomorphic
to $\Gamma_{n,n}\oplus X$. This is not a coincidence. Indeed, it is
known from the general theory of lattices (see Theorem A2
in the Appendix A), that for the 
sublattice in the unimodular lattice, its discriminant-form
is minus the discriminant-form of the orthogonal sublattice. Since
for our lattices the discriminant-forms take values in half-integers,
this means that the discriminant form of the sublattice is equal to
the discriminant-form of the orthogonal lattice. Since that rank and the
order of $X^*/X$ for our lattice $X$ satisfies the conditions of the 
Theorem A1 from Appendix A, the lattice $X$ is
uniquely determined by its rank, signature and the discriminant-form.}
$$
(D_4\oplus D_4 )^{\perp}=\Gamma_{3,3}\oplus D_4\oplus D_4
$$
--- this is our result for the seven-dimensional lattice. 
The symmetries of this lattice are the symmetries of
the M Theory background. Thus, the dual theory has the same moduli space
as the one we have obtained for the seven-dimensional CHL string.

\newsec{Cusps in the Moduli Space.}
One characteristic of the duality group is how many 
cusps we have in the moduli space of the theory. 
The cusps may be considered as the possible ways for going far away 
in the moduli space.  As we have explained in the section 4,
the point of the moduli space may be represented by the 
$10-D$-dimensional plane
in $\IR^{10-D,18-D}$, such that the scalar product is negative
definite on it. The moduli space is not compact, and the infinities
correspond to the possible degenerations of this scalar product.
If there is at least one null-vector on the plane, then this
plane is infinitely far away. 

We will restrict ourselves with such degenerate planes only, that
 the maximal isotropic subspace is rational 
(the word ``rational'' means, that the
plane goes through sufficiently many points of the lattice:
one can introduce a basis, consisting of the lattice points).
The rational isotropic planes of dimension $p=1,\ldots 10-D$ parametrize
the rational components of the boundary of the moduli space:
these components consist of the $10-D-p$-dimensional planes
in the orthogonal complement to the given isotropic plane.
(The boundary components are themselves the Grassmanians
$Gr(10-D-p,\IR^{10-D-p,18-D-p})$). There is known in mathematical
literature a construction of compactifications of
the quotients of Grassmanians by the discrete group 
(in fact, arbitrary quotients of the form 
$\Gamma\backslash G/K$) whose boundaries consist of the rational
components only --- see \rCasselman\  for a recent discussion,
and references therein. On the other hand, it was argued recently 
\rDouglasHull, that compactifications with irrational boundary components
are physically meaningful and may be described in terms of noncommutative
geometry.

Since an arbitrary isotropic plane may be embedded into some
maximal isotropic plane, we will classify first the maximal 
rational isotropic planes in $\IR\otimes \Gamma_{(D)}$.

{\bf D=9.} There is only one light-like vector in
$\gii\oplus\ge$, modulo the symmetry group \rConway. This may be
explained in the following way. The group $\Aut(II_{1,9})$
acts on the space $\IR^{1,9}$, and the fundamental domain
for this group may be constructed. The boundaries of the
fundamental domain are the hyperplanes orthogonal to the 
simple roots. The Dynkin diagram for the system of simple roots 
is shown on Fig.2. One can prove by direct computation that the only one 
light-like vector on the boundary of the fundamental domain 
is minus the imaginary root of $\widehat{E_8}$ (the combination 
of roots of $\widehat{E_8}$, which has zero length square). 

Another proof is given in the footnote in the Section 3.3.

{\bf D=8.} In the lattice corresponding to the eight-dimensional theory
there are at least two rational light-like planes. One is
the standard light-like plane in the lattice
\eqn\firstcusp{
\Gamma_{2,2}\oplus D_8
}
and the other --- the standard light-like plane in
\eqn\secondcusp{
\gii(2)\oplus\gii\oplus\ge
}
We call ``standard'' the light-like plane, generated by the vectors
of the form
$$
\vtrd{0}{*}{0}{*}{0}
$$
(the notations are as explained in Section 3.3 after the formula 
\lightlike.)

Let us prove that each light-like plane is equivalent to 
one of these two.

Consider a rational light-like 2-plane 
$\omega\subset\gii\oplus\gii(2)\oplus\ge$. We will use a trick
which we have learned from  \rAspinwall. Consider the intersection
$\omega^{(1)}$ of $\omega$ with the subspace $v^{\perp}$, where $v$ 
is a standard light-like vector in $\gii(2)$.  
Then, we can apply such a symmetry of 
$\gii\oplus\ge$, that the projection of $\omega^{(1)}$
onto $\gii\oplus\ge=v^{\perp}/v$ is the standard light-like line in
$\gii\oplus\ge$. This means that $\omega^{(1)}$ is now the rational
light-like line in the standard light-like plane in $\gii(2)\oplus\gii$.
Using the Lemma from Appendix C, we may bring it ot one
of the two standard forms: either with the orthogonal complement
in $\gii(2)\oplus\gii$ being $\gii$, or with the orthogonal
complement $\gii(2)$.  The other rational vector $v'$, generating the
light-like plane $\omega$ may be chosen to belong to $\gii\oplus\ge$
in the first case, or to $\gii(2)\oplus\ge$ in the second case. 
In the first case, we can apply the symmetry of $\gii\oplus\ge$ which
brings $v'$ to the standard form in $\gii\oplus\ge$, thus we get
$\omega$ the standard light-like plane in \secondcusp. In the second case,
$v'$ may be brought to one of the two standard forms into $\gii(2)\oplus\ge$,
as described in the Appendix C, which gives us either the case 
\firstcusp, or the case \secondcusp. 

{\bf D=7.} Now consider compactification to seven dimensions. 
The seven-dimensional lattice may be viewed in the three ways:
\eqn\sdl{
\gii\oplus\Gamma_{2,2}(2)\oplus\Gamma_8\simeq
\Gamma_{2,2}\oplus\gii(2)\oplus D_8\simeq
\Gamma_{3,3}\oplus D_4\oplus D_4
}
The standard lightlike planes in each of these three lattices
give three non-equivalent cusps.
Let us prove that there are no other cusps.

Consider a primitive lightlike vector $v\in\gii(2)$. Consider the
projection of $\omega'=\omega\cap v^{\perp}$ to 
$v^{\perp}/v\simeq\gii\oplus\gii(2)\oplus\ge$. 
This is a 
light-like 2-plane, and it is equivalent to one of the two
standard planes, as discussed above. Thus, we have two cases to
consider:

{\bf The first case:} 
The projection is equivalent to the standard light-like plane in
$\gii\oplus\gii(2)\oplus\ge$. Then, by the Lemma from 
Appendix C, the orthogonal complement to $\omega'$ in 
$\gii\oplus\Gamma_{2,2}(2)$ is either  $\gii$, or $\gii(2)$.
In case if it is $\gii$, our plane $\omega$ is generated
by the standard light-like plane in $\Gamma_{2,2}(2)$ and
some light-like vector in $\gii\oplus\ge$, which 
by the symmetry of $\gii\oplus\ge$
may be brought to the unique standard form. Thus, in
this case $\omega$ is equivalent to the standard plane in the
first lattice in \sdl. In the case if the orthogonal complement
is $\gii(2)$, we have $\omega$ generated by the standard plane
in $\gii\oplus\gii(2)$ and some lightlike vector 
$v'\in\gii(2)\oplus\ge$. Since this vector $v'$ is equivalent to
one of the two standard forms, we get either the standard plane
in the first, or in the second lattice in \sdl.

{\bf The second case:}
The projection is equivalent to the standard light-like plane in
$\Gamma_{2,2}\oplus D_8$. Then, by the Lemma from Appendix C,
the orthogonal complement to this plane in $\Gamma_{2,2}\oplus\gii(2)$
is either $\gii$, or $\gii(2)$. If it is $\gii$, then the vector
$v'$ in $\omega$ may be chosen to belong to $\gii\oplus D_8$. There
are two standard light-like vectors in $\gii\oplus D_8$, thus 
we get the standard plane in the second or the third lattice in
\sdl. If it is $\gii(2)$, then $v'$ belongs to $\gii(2)\oplus D_8$,
and the Lemma from Appendix C gives two possibilities, corresponding
to the standard light-like planes in the second or the
third lattices in \sdl.

In the case of partial decompactification, the plane 
contains one or more rational light-like vectors, 
which then may be included into one of the standard lightlike planes.
Then, the Lemma from Appendix D may be applied to bring
these vectors to one of the two standard forms 
within the standard plane.

We should explain the physical meaning
of these cusps. 

Let us prove that
the points far away in the moduli space correspond to
the degeneration of the metric on the torus. 
The $d+1$ -dimensional plane corresponding to the point of
the moduli space with the Wilson lines $A_i=(a_i,a_i)$,
the metric $g_{ij}$ and the 2-form $b_{ij}$ may be specified as
the plane orthogonal to the $d+9$ -dimensional plane with
the equation
\eqn\orthplane{
G_{ij}l^j=n_i+(a_i\cdot R)
}
where
\eqn\Gij{
G_{ij}=g_{ij}+b_{ij}+{1\over 2}(a_i\cdot a_j)
}
This implies, that in the plane corresponding
to our background we may pick a basis of vectors $\{X_p\}$, given
by the following formula:
\eqn\Xp{
X_p=\vtr{a_p}{\delta_p^i}{-G_{ip}}
}
There is a subgroup of the duality group, which acts on the background
by shifts
$$
a_i\to a_i-k_i\alpha,\;\;\; G_{ij}\to G_{ij}-(a_i\cdot\alpha)k_j+k_ik_j
$$
where $\alpha$ is a root of $E_8$ and $\{k_i\}$ is the set of integers.
This means, that the point at infinity in the moduli space may be 
attained with the Wilson lines kept finite. Also, since $b_{ij}$ is defined 
modulo a shift by an integer antisymmetric matrix, we may 
assume that $b_{ij}$ has finite limit. 
The  matrix of scalar products for the basis \Xp\ is:
\eqn\Gramm{
(X_p\cdot X_q)=-2g_{pq}
}
This means that the plane generated by $\{X_p\}$ is at finite distance
unless the metric $g_{pq}$ becomes singular. Thus, the infinities
of the moduli space correspond to the degenerations of the metric
on the torus. Degeneration of the metric means that at least one
eigenvalue of $g_{ij}$ goes either to infinity, or to zero.
If the corresponding eigenvector has rational coordinates, then
we may interpret this limit as shrinking some cycle
on a torus or considering a cycle of very large size.

If we think of the cusps as the possible
ways for going to infinity in the moduli space, then different
light-like planes are physically inequivalent. But if we consider
the limiting theories, then some identifications occur.
Indeed, notice that all the lightlike planes found have
a primitive vector from some $\Gamma_{1,1}$ sublattice in them.
This means, that the corresponding limiting theories
may be obtained by decompactification of the ninth dimension,
giving a usual heterotic string. In the moduli space of
the heterotic string on a torus, we have two cusps. They correspond
to the ten-dimensional $E_8\times E_8$ and $Spin(32)/\IZ_2$
heterotic string. In fact, the plane with the orthogonal
complement $E_8$ in $\Gamma_{2,2}\oplus D_8$ 
gives $E_8\times E_8$ theory,
while the plane with orthogonal complement $D_8$ gives 
$Spin(32)/\IZ_2$ theory. In the seven-dimensional lattice
$\Gamma_{3,3}\oplus D_4\oplus D_4$, the plane with the orthogonal
complement $E_8$ gives $E_8\times E_8$ heterotic string in ten
dimensions, and those whose complements are $D_8$ or $D_4\oplus D_4$
give $Spin(32)/\IZ_2$ heterotic string. To prove this, one has
to look at the background of the heterotic string corresponding
to the given cusp of the CHL string after decompactification
of the ninth dimension. 

\bigbreak\bigskip\bigskip\centerline{{\bf Acknowledgments}}\nobreak

I would like to thank J.H.~Conway, A.~Losev, D.R.~Morrison, 
S.~Sethi, E.~Sharpe and especially E.~Witten for many interesting 
and helpful discussions.  This work was supported in part by RFFI
Grant No. 96-02-19086 and partially by grant 96-15-96455 for
support of scientific schools.

\appendix{A}{Discriminant-form.}
There is a very convenient invariant of the lattices, called
the discriminant-form \refs{\rNikulin}.
Suppose we  are given an integer lattice
$L$ (the scalar product of any two vectors is integer). 
Let us consider the abelian group:
$$
A_L=L^*/L
$$
Since the scalar product on the lattice is integer valued,
we have a well-defined scalar product $b$ on $L^*/L$ taking
values in ${\bf Q}/\IZ$:
$$
b:A_L\times A_L\to {\bf Q}/\IZ
$$
Also, if  $L$ is an even lattice, the quadratic form on
the lattice $L$ defines a ${\bf Q}/2\IZ$-valued quadratic
form on $A_L$:
$$
q:A_L\to {\bf Q}/2\IZ
$$
The pair $q_L=(q,b)$ has a property
$$
q(a+a')-q(a)-q(a')\equiv 2b(a,a')\;\; \mod 2\IZ
$$
The finite abelian group $A_L$ together with the pair $q_L=(q,b)$
is an invariant of the lattice, called the discriminant-form.
It turns out, that in many 
important cases this invariant, together with
the signature of the lattice,  characterizes the lattice 
unambiguously. Namely, there is the following
theorem by Kneser and Nikulin\foot{We use 
the version of this theorem, formulated in \rMorrison\ (Theorems
2.2 and 2.8).}:

{\bf Theorem A1.} {\it Let $L$ be an even lattice with the 
the signature $(s_+,s_-)$, $s_+>0$, $s_->0$, and the minimum 
number $l(A_L)$ 
of generators of $A_L$ satisfies an inequality:
$$l(A_L)\leq {\rm rank}(L)-2$$
Then there is only one even lattice $L$ with invariants
$(s_+,s_-,q_L)$. 

If $M$ is an even lattice with invariants
$(t_+,t_-,q_M)$ and $L$ is an even unimodular lattice
of signature $(s_+,s_-)$, subject to inequalities
$$
\matrix{t_+<s_+,\cr t_-<s_-,\cr l(A_M)\leq {\rm rank}(L)-
{\rm rank} (M) - 2}
$$
then there is a unique 
primitive\foot{Primitive means that primitive vectors
of $M$ go to primitive vectors of $L$}
embedding of $M$ into $L$.}

The reader should consult \refs{\rMorrison,\rNikulin}
for further details and references.  

Let us compute the discriminant-forms for the CHL momentum
lattices in various dimensions. 

1) $L=\Gamma_{n,n}(2)^*/\Gamma_{n,n}(2)$: 
In this case $A_L=(\IZ_2\oplus \IZ_2)^n$,
generated by $2n$ elements $x_i=(\half,0)$
and $y_i=(0,1)$ in the $i$-th factor
$\gii(2)_{(i)}^*\simeq\gii(\half)_{(i)}$. The quadratic form
and scalar product are $q(x_i)\equiv q(y_i)\equiv 0$ and 
$b(x_i,y_j)\equiv \half\delta_{ij}$. 

2) $L=D_8$: $A_L=D_8^*/D_8=\IZ_2\oplus \IZ_2$, generated
by $v=(1,0^{(7)})$ and $s=\left({\half}^{(8)}\right)$. 
The invariants are $q(v)\equiv 1$, $q(s)\equiv 0$,
$b(v,s)\equiv \half$. The transformation $x=v+s$, $y=s$
is an isomorphism between $A$ for $D_8^*/D_8$ and
$A$ for $\gii(2)^*/\gii(2)$. Indeed, we have:
$q(x)\equiv q(y)\equiv 0$ and $b(x,y)\equiv \half$. 
Thus,  the theorem by Nikulin and Kneser implies:
$$
\gii(2)\oplus \ge\simeq \gii\oplus D_8
$$

3) $L=D_4\oplus D_4$: $A_L=(\IZ_2\oplus \IZ_2)^{\oplus 2}$, 
$q(v_i)\equiv q(s_i)\equiv 1$, $b(v_i,s_j)\equiv\half\delta_{ij}$, 
$i=1,2$. This is isomorphic to $\Gamma_{2,2}(2)^*/\Gamma_{2,2}(2)$:
$x_1=v_1+v_2,\;\; x_2=s_1+s_2,\;\;
y_1=v_1+v_2+s_1,\;\; y_2=s_1+s_2+v_1$. This tells us that
$$
\Gamma_{2,2}(2)\oplus \ge\simeq \Gamma_{2,2}\oplus D_4\oplus D_4
$$

4) $L=D_8^*(2)$: $A_L$ is generated by six elements 
of the form $a_j=\stwi(u_j+u_{j+1})$ for $j=1,\ldots, 6$.
$q(a_j)\equiv 1$, $b(a_i,a_{i+1})\equiv\half$. An isomorphism with
$A_L$ for $L=\Gamma_{3,3}(2)$ is established as follows
(shown are the images of the generators of 
$\left[D_8^*(2)\right]^*/D_8^*(2)$
in $\Gamma_{3,3}(2)^*/\Gamma_{3,3}(2)$):
$$
\matrix{(a_1,a_2,a_3,a_4,a_5,a_6)=\cr
\cr
\left(
\left[\matrix{\half&1\cr \half&0\cr\half&0}\right],\;
\left[\matrix{\half & 1\cr 0&1\cr 0&0}\right],\;
\left[\matrix{0&1\cr 0&0\cr \half& 1}\right],\;
\left[\matrix{\half&1\cr 0&0\cr 0&0}\right],\;
\left[\matrix{0&1\cr\half&1\cr 0&0}\right],\;
\left[\matrix{0&0\cr 0&1\cr \half&1}\right]
\right)
}
$$
This isomorphism implies
$$
\Gamma_{3,3}(2)\oplus \Gamma_8\simeq \Gamma_{3,3}\oplus D_8^*(2)
$$

In Section 5 we needed the following property of the Nikulin lattice
$\nik=D_8^*(-2)$: the projection map $O(\nik)\to O(A_{\nik})$
is surjective. We will prove it, using the method suggested in
Section 14.1 of \rNikulin. Let us consider the primitive embedding
$f:\nik\to \Gamma_{16}$. We will need the following theorem

{\bf Theorem A2.} (Proposition 1.6.1 in \rNikulin.)
{\it A primitive embedding of an even lattice $S$ into an even 
unimodular lattice $L$, in which the orthogonal complement is isomorphic
to $K$, is determined by an isomorphism $\gamma: 
A_S\simeq A_K$ for which 
\eqn\Ngamma{q_K\circ \gamma= -q_S}
Two such isomorphism $\gamma$ and $\gamma'$ determine isomorphic
primitive embeddings if and only if they are conjugate via an isomorphism
$g$ of $K$: $\gamma=\bar{g}\gamma'$ where $\bar{g}$ is the action
of $g$ on $A_K$.}

(Two embeddings $i_1:S\to L$ and $i_2:S\to L$ are considered as
isomorphic, if there is an isomorphism $h$ of $L$ with the property
$i_2=hi_1$.) 

One can prove by direct computation, that
there is a unique up to isomorphism of $\Gamma_{16}$
embedding of $\nik$ into $\Gamma_{16}$, whose orthogonal
complement is $\nik^{\perp} \simeq\nik$ (this embedding
may be constructed as follows: $\nik$ is generated by
$E_1,\ldots,E_{8}$ with $E_i\cdot E_j=2\delta_{ij}$
and $\half(E_1+\cdots+E_8)$. Then, $f(E_i)=u_{2i-1}+u_{2i}$.
To prove that it is a unique embedding, one has to take into
account that vectors $E_i\in\nik$ should go 
to the roots of $\Gamma_{16}$, which are of the form $\pm u_i\pm'u_j$.
Then, one has to take into account that 
$\half\sum\limits_{i=1}^8 E_i\in\nik$.)
Thus, for arbitrary $h\in O(A_{\nik})$ there is
such $g\in O(\nik)$ that $h \bar{g}={\rm id}$. 

In the same way, one can prove that the map 
$O(\Gamma)\to O(A_{\Gamma})$ is surjective for 
$\Gamma=D_8$ (by considering the unique primitive embedding
$D_8\subset \Gamma_{16}$) and $\Gamma=D_4\oplus D_4$
(which may be embedded into $\ge\oplus\ge$ in only one way).
For the lattice $\Gamma_8(2)$ we will prove it in a
different way. We have $A_{\ge(2)}\simeq \ge\left(\half\right)/\ge(2)$.
Let us denote $\pi: \ge\left(\half\right)\to A_{\ge(2)}$ the
natural projection.
The group $\ge/2\ge$ is generated by the projections $\pi(\alpha_i)$ 
of the simple roots of $\ge$. Given an automorphism $\bar{g}$ 
of this group which preserves the discriminant form, we consider the 
images $\bar{g}(\pi(\alpha_i))$ of the projections of the 
simple roots of $\Gamma_8$. Notice
that $q(\bar{g}(\alpha_i))\equiv 1\mod 2\IZ$. As we have explained
in Section 3.1, an arbitrary element of $\ge\left(\half\right)$,
whose length square is odd, is equivalent modulo $\ge(2)$ to 
the element with length square one. This means, that we may
choose representatives of $\bar{g}(\pi(\alpha_i))$ being 
$\stwi \beta_i$, where $\beta_i$ is some root of $\ge$.
Since $\bar{g}$ preserves the discriminant-form, the scalar
products of the roots $\beta_j$ have the following form:
$$
(\beta_i\cdot\beta_j)=\left\{\matrix{2\;\; {\rm if}\;\; i=j\cr
\pm 1 \;\; {\rm if}\;\; i\neq j\;\; {\rm and}\;\; 
(\alpha_i,\alpha_j)=-1\cr
0\;\; {\rm in\;\; other \;\; cases}
}\right.
$$
Now take into account that we have a freedom to change a sign of
$\beta_j$ (because $\beta_j\equiv -\beta_j\mod 2\ge$). Since
the Dynkin diagram of $\ge$ does not contain cycles, we may
adjust signs of the roots $\beta_j$ in such a way that they
form a system of simple roots of $\ge$. But it is known from
the theory of Lie groups (\rVinberg, p.78), 
that any two systems of simple roots are related by 
some symmetry $g$ of the lattice. This symmetry is the required lift
of $\bar{g}$.

As another example of how Theorem A2 works, let us prove the
isomorphism which we used in Section 4:
\eqn\isosecfour{
\Gamma_{n,n}\simeq \{(w;\tilde{w})|w,\tilde{w}\in D_n^*,
\; w-\tilde{w}\in D_n\}
}

There is an evident map $g: D_n(1)\to D_n(-1)$  which multiplies
scalar product by $-1$. The associated map  $\bar{g}: A_{D_n(1)}\to
A_{D_n(-1)}$ satisfies \Ngamma. Thus, we may glue 
$D_n(1)$ and $D_n(-1)$ into an even unimodular lattice,
by adding to $D_n(1)\oplus D_n(-1)$ vectors of the form
$(v,gv)$ with $v\in D^*_n(1)$. In other words, we consider
such vectors $(x,y)\in D^*_8(1)\oplus D_8^*(-1)$ that
the conjugacy class of $x$ coincides with the conjugacy class
of $y$. The lattice generated by these vectors is even and self-dual,
and it coincides with $\Gamma_{n,n}$.

Also, we used in Section 5 the fact that the lattice 
$\Gamma_{1,1}\oplus\Gamma_{n-1,n-1}(2)$ is isomorphic
to the lattice generated by $(\stw w,\stw \tilde{w})$ with
$w$ and $\tilde{w}$ weights of $D_n$, and $w-\tilde{w}$
is either in the root lattice, or in the vector conjugacy
class of $D_n$.
Let us prove it. For convenience, we divide the scalar
product by two. Consider the lattice 
$\gii\left(\half\right)\oplus\Gamma_{n-1,n-1}$. 
This lattice may be obtained from the lattice $\Gamma_{n,n}$
by adding one half of the primitive lightlike vector (it
follows from Appendix D, that there is only one primitive
lightlike vector in $\Gamma_{n,n}$, modulo the symmetries).
We know, that $\Gamma_{n,n}$ is isomorphic
to the lattice $\{(w;\tilde{w})|w-\tilde{w}\in D_8\}$.
The vector $(2s,2\bar{s})$, where $s$ and $\bar{s}$ are
spinor and conjugate spinor of $D_n$, is primitive and lightlike.
Adding the generator $(s,\tilde{s})$ we get the lattice consisting of
$(w,\tilde{w})$ with $w-\tilde{w}$ is in scalar or vector
conjugacy class of $D_n$. This is what we had to prove.
 
\appendix{B}{Example of $K3$ with Nikulin involution.}
Consider the surface $\bf X$ in the weighted projective space $\IP_{6,4,1,1}$, 
specified by the following equation:
\eqn\Surface{
y^2=x(x^2+a(u,v)x+b(u,v))
}
where $a(u,v)$ and $b(u,v)$ are homogeneous polynomials of degree 
4 and 8. We will work in the affine coordinates with $v=1$,
and denote $a(u)=a(u,1)$, $b(u)=b(u,1)$.
Notice that we have eight $A_1$ singularities, located at the 
points $x=y=0$, $b(u)=0$. Indeed, our equation \Surface\ 
near such a singularity may be approximated as:
\eqn\NearSing{
y^2=ax(x-c(u-u_0))
}
where ${b\over a}=c(u-u_0)+\ldots$. This is the equation for $A_1$ 
singularity, located at $x=y=0$, $u=u_0$.
The existence of these eight singular points is not an accident.
They may be interpreted as singularities of the quotient
of another $K3$ surface by $\IZ_2$-involution.

Let us explain it.
The surface \Surface\ admits an involution 
$x\to \tilde{x}$, $y\to\tilde{y}$
where $\tilde{x}$ and $\tilde{y}$ are given  by the following equations:
\eqn\tildas{
\matrix{ x\tilde{x}=b(u)\cr {\tilde{y}\over\tilde{x}}=-{y\over x}
}}
This involution preserves the holomorphic 2-form 
\eqn\HolForm{
\Omega={du\wedge dx\over y}
}
and has eight fixed points located at $y=0$, $x={a\over 2}$,
$D=a^2-4b=0$. Let us consider the quotient by this involution.
The invariant functions are:
\eqn\invariants{
u,\;\;Y={y\over x}\left(x-{b\over x}\right),\;\;
X=\left({y\over x}\right)^2=x+{b\over x}+a
}
They satisfy the equation
\eqn\NewSurface{
Y^2=X(X^2-2aX+(a^2-4b))
}
This is, again, an elliptic K3 surface with a double section.
One can see that there is a reciprocal relation 
between the surfaces \Surface\ and \NewSurface: one is the
$\IZ_2$-quotient of the other, and the singularities of one
surface correspond to the fixed points of $\IZ_2$-involution of the other.

In this situation, as we have mentioned in Section 5, the lattice
generated by the exceptional divisors $E_1,\ldots,E_8$ contains
$\half\sum\limits_{k=1}^8 E_k$. The appearence of the half of the
sum of exceptional divisors may be explained as follows. 
Suppose that we have the surface $\bf X$, with the involution 
$\iota$, and ${\bf Y}={\bf X}/\iota$. There are many meromorphic
functions on $\bf X$, anti-invariant under $\iota$. (One can
take an arbitrary meromorphic function $\psi$, not invariant under
$\iota$, and consider $\phi=\psi-\iota^*\psi$ --- this function will
be anti-invariant.) The general anti-invariant function $\phi$ will
have first order zeroes at each of the eight fixed points of $\iota$. 
Going to the quotient ${\bf Y}={\bf X}/\iota$, we may consider
this anti-invariant function as a two-valued function on ${\bf Y}$: 
when we circle around the exceptional divisor $E_i$, this function 
changes its sign. Let us call this two-valued function $\phi_{\bf Y}$.
If the exceptional divisor is given locally
by the equation $\chi(X,Y,u)=0$, then the function $\phi_{\bf Y}$
locally near the exceptional divisor may be written as 
$\phi_{\bf Y}\sim\sqrt{\chi}$ --- in other words, it has zero
of order $\half$ at each $E_i$. Also, the function $\phi_{\bf Y}$
may have zeroes and poles of integer order at some other
surfaces on $\bf Y$. Notice that $\phi_{\bf Y}^2$ is a well-defined
(single-valued) function on $\bf Y$. The corresponding principal
divisor is:
\eqn\DivPhy{
(\phi_{\bf Y}^2)=\sum\limits_{j=1}^8 E_j + 2 v
}
where $v$ is some divisor (combination of cycles with integer coefficients)
on $\bf Y$. This means, that the sum of the exceptional
divisors is equivalent to the cycle $-2v$ in 
homology\foot{Linear equivalence of divisors implies topological
equivalence: given $(f)=(f)_0-(f)_{\infty}$, 
the cycle given by the equation $f=t$ interpolates between the cycle 
$f=0$ and the cycle $f=\infty$, as $t$ runs from $0$ to $\infty$.} 
--- in other 
words, the cycle $\sum\limits_{j=1}^8 E_j$ may be divided by two.
The converse is also true. Consider the K3 manifold $Y$ with eight
$A_1$ singularities, such that half of their sum belongs to the
Picard lattice of $Y$. This means that the sum of exceptional
divisors may be expressed as:
\eqn\Converse{
\sum\limits_{j=1}^8 E_j=2\sum n_j c_j
}
where $c_j$ are some holomorphic curves in $Y$, $n_j\in\IZ$.
Since for the holomorphic curves on K3, the topological equivalence
implies linear equivalence (\rIS, Ch.12), there exists a meromorphic
function with first order zero at each exceptional divisor,
and even order zeroes or poles at curves $c_j$. The square root
$\phi_{\bf Y}$ of this function is a two-valued function on $Y$,
ramified at the exceptional divisors. The graph of $\phi_{\bf Y}$
(surface in $Y\times\IP^1$) is some surface $X$. When the exceptional
divisors $E_j$ are blown down, this surface has a well-defined 
2-form.  

Let us make this reasoning more explicit for the particular 
example of the surface \Surface\ and its $\IZ_2$-quotient \NewSurface.
The Picard group of the surface \NewSurface\ is generated by the
Poincare duals of the following holomorphic cycles:
\eqn\HolCycles{\matrix{
s:\;\;&(X=Y=0) \;\;& {\rm -\; the \; section},\cr
\tilde{s}:\;\;&(X=Y=\infty)\;\;& {\rm -\; the \; other \; section},\cr
\xi:\;\; &(X^2+aX+b=0,\;Y=0)\;\;& {\rm -\; the \; double \; section},\cr
f: \;\; &(u={\rm const})\;\;& {\rm - \; the \; fiber},
}}
and the eight exceptional divisors $E_i,\;i=1,\ldots,8$. 
Consider the function 
$$\phi(x,y,u)={x-{b\over x}\over x+{b\over x}+a}$$ 
on the surface \Surface. This function is anti-invariant. We have:
\eqn\PhiY{
\phi_{\bf Y}^2(X,Y,u)={Y^2\over X^3}
}
--- the function on the surface \NewSurface. This function has 
the second order zero at the double section $\xi$, and the fourth
order pole at the section $s$. It also has the first order 
pole at each $E_i$. Indeed, the equation \NewSurface\ in the vicinity
of $E_i$ may be rewritten as
\eqn\NearE{
Y^2\simeq -2aX(X-\alpha (u-u_0))
}
--- this is the equation for the $\IZ_2$ singularity. To blow
up, we have to glue in the sphere $\IP^1$, so that 
\NearE\ becomes ${\cal O}_{\IP^1}(-2)$. This is equivalent to
introducing a function $z$ (coordinate on $\IP^1$), such that
$z^2={X-\alpha (u-u_0)\over X}$. Then, our function $\phi_{\bf Y}^2$
becomes ${Y^2\over X^3}\sim {z^2\over X}$ which means that it has
the first order pole at the divisor. This proves that the divisor
$$
2\xi-4s-\sum\limits_{j=1}^8 E_j
$$
is principal, thus it is topologically trivial. In other words,
\eqn\queer{
\half\sum\limits_{j=1}^8 E_j=-2s+\xi
}
Also, by looking at poles and zeroes of the function $Y\over X$,
we learn that 
\eqn\another{
2f+s+\tilde{s}-\xi=0
}
Notice that the holomorphic cycles \HolCycles\ with the
relations \queer\ and \another\ generate the Picard lattice
$\Gamma_{1,1}\oplus{\cal N}$.

\appendix{C}{Lightlike lines in some Lorentzian lattices.}
In this section, we will classify the lightlike lines in two
Lorentzian lattices: $\gii\oplus D_8$ and $\gii\oplus D_4\oplus D_4$.
We will actually give a proof for the second lattice only, 
since the first one can be considered in the very
similar way. We may represent this lattice in two forms:
\eqn\GDD{
\gii\oplus D_4\oplus D_4\simeq \gii(2)\oplus D_8}

This gives two inequivalent examples of light-like lines:
the standard light-like line in the first lattice and the one
in the second lattice. They cannot be related by the automorphism
of the lattice, since they have different orthogonal complements. 
Are there any lightlike lines, which are not equivalent to 
one of these two?

Consider some light-like primitive vector:
\eqn\llpv{
\left[\matrix{P_1&P_2\cr m&-n}\right]
}
where $P_1$ and $P_2$ are some vectors from $D_4$. Our strategy
will be to try to decrease $\max(P_1^2,P_2^2)$ by acting on this
vector with some symmetry of the lattice.
Without any loss of generality, we may suppose that 
$P_1^2\geq P_2^2$ and $n\geq m>0$. Let us consider the following 
symmetry:
\eqn\sym{\left[
\matrix{P_1 & P_2\cr m&-n}\right]\to
\left[\matrix{P_1-m\alpha & P_2\cr m & -m-n+(P_1\cdot\alpha)}\right]
}
with $\alpha^2=2$ and $(\alpha\cdot P_1)\geq 0$.
The condition that $(P_1-m\alpha)^2\leq P_1^2$ may be written as:
\eqn\aPgm{
(\alpha\cdot P_1)\geq m
}

From $P_1^2+P_2^2-2mn=0$ we infer that $m\leq\sqrt{P_1^2}$. 
(This inequality is saturated if and only if $P_1^2=P_2^2$
and $m=n$.) Let us prove that for most of the values of $P_1$
we can find $\alpha$ which satisfies the inequality not weaker
then \aPgm:
$$
(\alpha\cdot P_1)\geq\sqrt{P_1^2}
$$

Indeed, suppose $P_1=(p_1,p_2,p_3,p_4)$ and 
$p_1\geq p_2\geq p_3\geq p_4\geq 0$ (we may always
bring $P_1$ to this form by some symmetry of $D_4$).
Then, let us take $\alpha=(1,1,0,0)$. This choice
gives 
$$
(p_1+p_2)\geq\sqrt{p_1^2+p_2^2+p_3^2+p_4^2}
$$
with equality if and only if $p_1=p_2=p_3=p_4=p$
or $p_2=p_3=p_4=0$. Thus, we may
continue this process until we reach $P_1=(p,p,p,p)$ or $(2p,0,0,0)$ 
and $m=n$. At this point we should have $P_2^2=P_1^2=4p^2$ and $m=n=2p$.
If now $P_2$ is not of the form $(p,p,p,p)$ or $(2p,0,0,0)$, then we
can continue to play \sym, but this time subtracting $m\alpha$ from
$P_2$, not from $P_1$. This process stops when we get the vector of 
one of the four types:
\eqn\fourtypes{
\left[\matrix{0&0\cr 1&0}\right]\;\;,\;\;
\left[\matrix{(1111)&(1111)\cr 2&-2}\right]\;\;,\;\;
\left[\matrix{(2000)&(1111)\cr 2&-2}\right]\;\;,\;\;
\left[\matrix{(2000)&(2000)\cr 2&-2}\right]
}
but the third and the fourth are related to the second via 
triality of $D_4$ (the outer automorphism). Thus, we have essentially
only two possibilities. This proves that there are only two lightlike
lines, given by the standard lightlike lines
in \GDD\ (the first one in \fourtypes\ has orthogonal complement
$D_4\oplus D_4$, and the second one $D_8$). 

Similar (but simpler) reasoning for $\gii\oplus D_8$ shows
that there are two and only two light-like lines there, which may
be thought of the standard in two representations of the lattice:
$$
\gii\oplus D_8\simeq \gii(2)\oplus E_8
$$

\appendix{D}{Technical Lemma.}
Consider a standard light-like plane in
$\Gamma_{n,n}\oplus\Gamma_{m,m}(2)$, and a primitive vector $v$ in it.
Here we will prove that this vector may be mapped by the
symmetry of the lattice to one of the two standard forms: one
standard form being a standard light-like vector in the first $\Gamma_{1,1}$
sublattice, and the other the standard light-like vector in the first
$\gii(2)$ sublattice. 

To prove this, consider first the primitive vector $w$
in the standard light-like plane in the lattice $\Gamma_{2,2}$. 
It has the following form: 
\eqn\w{
w(x,y)=\left[\matrix{x&0\cr y&0}\right]
}
Let us prove that we may bring it by the symmetries of $\Gamma_{2,2}$
to the form
\eqn\wSt{
w({\rm GCD}(x,y),0)=\left[\matrix{{\rm GCD}(x,y)&0\cr 0&0}\right]
}
where ${\rm GCD}(x,y)$ is the greatest common divisor of $x$ and $y$.
Indeed, for an arbitrary pair $(p,q)$ of integers, we may consider the
following symmetries of the lattice $\Gamma_{2,2}$:
\eqn\Euclid{
\left[\matrix{a & c\cr b & d}\right]
\to\left[\matrix{a+pb & c\cr b & d-pc}\right]
\;\;\; {\rm and}\;\; 
\left[\matrix{a & c\cr b& d}\right]
\to\left[\matrix{a& c+qd\cr b-qa& d}\right]
}
These symmetries map $w(x,y)$ to $w(x+py,y)$ and $w(x,y-qx)$.
But the transformations $(x,y)\to (x+y,y)$ and $(x,y)\to (x,y-x)$
on the pair of integers are the basic transformations of the
Euclid algorithm for the greatest common divisor.
By these transformations, one can relate $(x,y)$ to $({\rm GCD}(x,y),0)$.

Applying these transformations to the components of $v$ in $\Gamma_{n,n}$
and $\Gamma_{m,m}(2)$, we can map it to the vector in 
$\gii\oplus\gii(2)$ --- the direct sum of the first $\gii$
sublattice and the first $\gii(2)$ sublattice. Suppose this vector
has a form:
\eqn\v{
v(a,b)=\left[\matrix{a&0\cr b&0}\right]
}
Then, we can apply the transformations \Euclid\ of the 
Euclid algorithm to the pair $(a,b)$, with the only restriction 
that $p$ should now be even (we realize the lattice $\gii\oplus\gii(2)$
as consisting of the vectors of the form 
$\left[\matrix{a&b\cr c&d}\right]$, with the
condition that $d$ is even). This allows us to bring $v(a,b)$
to one of the vectors $v(1,0)$ or $v(0,1)$ (we have taken into
account that our vector is primitive).  This proves the lemma.

In the similar way, one can prove that an arbitrary $p$-dimensional
rational subspace in the standard lightlike plane in the lattice
$\Gamma_{n,n}\oplus\Gamma_{m,m}(2)$ may be mapped to the standard
lightlike plane in some 
$\Gamma_{p_1,p_1}\oplus\Gamma_{p_2,p_2}$ --- the direct sum of
the first $p_1$ $\gii$ sublattices and the first $p_2$ $\gii(2)$
sublattices, with the condition $p_1+p_2=p$.

\listrefs

\end